\documentclass[a4paper,12pt]{article}
\usepackage{epsfig}
\newcommand{\bmat}{\left(\begin{array}}
\newcommand{\emat}{\end{array}\right)}

\def\lsim{\raise0.3ex\hbox{$\;<$\kern-0.75em\raise-1.1ex\hbox{$\sim\;$}}}
\def\gsim{\raise0.3ex\hbox{$\;>$\kern-0.75em\raise-1.1ex\hbox{$\sim\;$}}}

\def\yzero{\smash{\hbox{$y\kern-4pt\raise1pt\hbox{${}^\circ$}$}}}

\def\s2{\frac{1}{\sqrt2}}

\def\beq{\begin{equation}}
\def\eeq{\end{equation}}
\def\beqa{\begin{eqnarray}}
\def\eeqa{\end{eqnarray}}

\def\IF{\relax{\rm I\kern-.18em F}}
\def\II{\relax{\rm I\kern-.18em I}}
\def\IP{\relax{\rm I\kern-.18em P}}
\def\IC{\relax\hbox{\kern.25em$\inbar\kern-.3em{\rm C}$}}
\def\IR{\relax{\rm I\kern-.18em R}}

\def\Dsl{\,\raise.15ex\hbox{/}\mkern-13.5mu D} 
\def\IZ{Z\kern-.4em  Z}
\def\bmat{\left(\begin{array}}
\def\emat{\end{array}\right)}

\def    \part          {\partial}
\def    \be            {\begin{equation}}
\def    \ee            {\end{equation}}
\def    \bea           {\begin{eqnarray}}
\def    \eea           {\end{eqnarray}}

\begin{document}
%
\pagestyle{empty}

\rightline{IPPP-02/100}
\rightline{FTUAM 02/34}
\rightline{IFT-UAM/CSIC-02-57}
\rightline{hep-ph/0212258}
\rightline{December 2002}

\renewcommand{\thefootnote}{\fnsymbol{footnote}}
\setcounter{footnote}{0}

\vspace{0.5cm}
\begin{center}
\large{\bf Quark and Lepton Masses and Mixing Angles from
Superstring Constructions
\\[5mm]}
\vspace{0.5cm}
\mbox{\sc{
\small{
S.A. Abel$^{1}$
}
}
}
{\small and}
\mbox{\sc{
\small{
C. Mu\~noz$^{2}$
}
}
}
\vspace{0.5cm}
\begin{center}
{\small
{\it 
$^1$
IPPP, Centre for Particle Theory,
Durham University, DH1 3LE, Durham, U.K. \\
\vspace*{2mm}
\it 
$^2$
Departamento de F\'{\i}sica
Te\'orica C-XI and
Instituto de  F\'{\i}sica
Te\'orica C-XVI,\\[-0.1cm]
Universidad Aut\'onoma de Madrid,
Cantoblanco, 28049 Madrid, Spain. \\
\vspace*{2mm}
} 
}
\end{center}

\vspace{1cm}

{\bf Abstract} 
\\[7mm]
\end{center}
\begin{center}
\begin{minipage}[h]{14.0cm}

We show that the observed structure of quark and lepton masses and
mixing angles can arise entirely geometrically from superstring 
constructions, at the renormalizable level. 
The model we consider is a $Z_3$ orbifold
compactification of heterotic string with two Wilson lines, where
three families of particles of $SU(3)_c\times SU(2)_L\times U(1)_Y$,
including Higgses,
are automatically present. In orbifold models, Yukawa couplings can
be calculated explicitly, and it is known that they get exponential
suppression factors depending on the distance between the fixed points
to which the fields are attached.  We find that in the $Z_3$ case, the
quark and charged-lepton mass hierarchies can easily be obtained for
reasonable values of the three moduli determining the radii of the
compactified space, $T_i\sim 1$.  For the neutrinos, due to the smallness
of their Dirac masses, the required scale for the see-saw mechanism 
to give the correct masses 
is found to be within reach of the electroweak scale. 
Finally, we find that one of the small number of possibilities for
quark and lepton mass matrices yields consistent results for the
mixing angles and the weak CP violation phase.  Although our scheme
relies on the mixing between fields due to
Fayet-Iliopoulos breaking, it is considerably more predictive than
alternative models of flavour.

\end{minipage}
\end{center}
%

%
\newpage
\setcounter{page}{1}
\pagestyle{plain}
\renewcommand{\thefootnote}{\arabic{footnote}}
\setcounter{footnote}{0}
%
%

\section{Introduction}

The Higgs mechanism \cite{Higgs} is the crucial ingredient in the
Standard Model required to explain electroweak symmetry breaking \cite{Weinberg}, 
and hence the masses of the $W^{\pm}$ and $Z$ gauge 
bosons. As an added bonus, 
the vacuum expectation value (VEV) of the Higgs field
generates fermion masses through Yukawa couplings.
However, the Standard Model does not address 
the origin of these couplings and the peculiar hierarchies
that are required to reproduce the observed structure of quark and
lepton masses \cite{pdg1}:
\bea
\ m_u  = 1.5\ to\ 4.5\ MeV\ ,\,\ \ m_c  = 1.0\ 
to\ 1.4\ GeV\ ,\,\ \ m_t=174.3\pm
5.1\ GeV 
\ ,
\nonumber
\\
\ m_d  =  5\ to\ 8.5\ MeV\ ,\,\ \ m_s=80\ to\ 155\ MeV\ ,\,\ \ m_b=4.0\ to\ 4.5\ GeV
\ ,
\nonumber
\\
\ m_e=0.51\ MeV\ ,\,\ \ m_{\mu}=105.658\ MeV\ ,\,\ \ m_{\tau}=1.777\ GeV
\ .
\label{masses}
\eea
In the framework of the Standard Model these are initial
parameters that are put in by hand.
In addition, the Yukawa couplings have to have off diagonal elements, 
with the Cabibbo--Kobayashi--Maskawa (CKM) weak coupling matrix 
\cite{Cabibbo} arising from the matrices that diagonalize the 
up- and down-quark mass matrices have the form \cite{pdg1}
\bea
V_{CKM}
=\left( \begin{array}{ccc}
0.9741\ to\ 0.9756 & 0.219\ to\ 0.226 & 0.0025\ to\ 0.0048\\
0.219\ to\ 0.226 & 0.9732\ to\ 0.9748 & 0.038\ to\ 0.044\\
0.004\ to\ 0.014 & 0.037\ to\ 0.044 & 0.9990\ to\ 0.9993
\end{array}\right) 
\ ,
\label{ckm}
\eea
%
The understanding, within some proposed extension of the Standard Model, of 
the particle masses (\ref{masses}) 
and the elements of the CKM matrix (\ref{ckm}),
remains one of the most important goals in particle physics.
In addition one would like to understand the mixings and hierarchies of 
neutrinos. The global analysis of solar, atmospheric and reactor
data \cite{Torrente} 
indicates that they have masses given by
%
%
\bea
\Delta m_{21}^2\approx 2.4\times 10^{-5}\ to\ 
2.4\times 10^{-4}\ eV^2
\ ,\,\ \ 
\Delta m_{32}^2\approx 
1.4\times 10^{-3}\ to\ 
6\times 10^{-3}\ eV^2
\ ,
\eea
and a 
Maki-Nakagawa-Sakata (MNS) weak coupling matrix \cite{mns} with
the charged leptons \cite{Torrente}:
\bea
V_{MNS}
=\left( \begin{array}{ccc}
0.73\ to\ 0.89 & 0.45\ to\ 0.66 & < 0.24\\
0.23\ to\ 0.66 & 0.24\ to\ 0.75 & 0.52\ to\ 0.87\\
0.06\ to\ 0.57 & 0.40\ to\ 0.82 & 0.48\ to\ 0.85
\end{array}\right) 
\ .
\eea
Since string theory is the prime candidate
for the fundamental theory of particle physics (from
which the Standard Model might be derived as a low-energy limit),
we think that it must be able, in principle, to tackle 
these questions directly. In this paper, we present a
stringy explanation for all of the fermion masses and mixings which 
is intended to take us a further step in this direction. Our approach will 
be `bottom-up', in the sense that we will not be presenting a completely 
explicit string construction, but will be asking if there is a particle 
assignment (to for example different twisted sectors) in a particular orbifold 
construction, that can explain the observed structure of masses.

String theory does provide some striking hints that 
a natural explanation for the structure of masses and mixings
might be possible. Indeed it is well known that
Abelian $Z_n$ orbifold compactifications \cite{Dixon,Wilson} 
of the Heterotic Superstring 
have a beautiful geometric mechanism to generate
a mass hierarchy \cite{Hamidi}--\cite{Faustino3}.
$Z_n$ orbifolds have twisted fields which are 
attached to orbifold fixed points. Fields at different fixed points 
can communicate with each other only via world sheet instantons.
The resulting renormalizable Yukawa couplings can be explicitly 
computed \cite{Hamidi},
\cite{Faustino}--\cite{Faustino2} 
and those between fields in twisted sectors 
get exponential suppression factors that depend on the
distance between the fixed points to which the relevant fields are
attached. 
These distances can be varied by giving different
VEVs to the $T$-moduli associated with the size and shape of the orbifold.
The question of hierarchies is then translated into the question of why 
the moduli fields take the values that they do. Although an explanation
for this lies outside the realm of perturbative string theory, generally 
there are far less moduli than there are hierarchies to be explained, and 
the nett outcome of such a geometric approach is an impressively large set of 
mass predictions\footnote{Note that throughout the paper we use the word 
``prediction'' in the very specific sense that it is commonly used when analyzing
fermion masses. That is one 
assumes values for a set of parameters (for example charges in the case of
Froggatt-Nielsen models). Then {\em given} those values one asks how many 
variables (such as VEVs) are fixed by masses. Once the variables are all fixed, 
the remaining masses are ``predictions''. In the present case, the $T$ moduli VEVs
are variables to be fixed by some of the masses, 
but the assumption of say a $Z_3$ orbifold on an 
orthogonal lattice can be considered as an initial assumption.}.

A strictly geometric explanation of Yukawa hierarchies means that the 
Yukawa couplings must be entirely renormalizable, since non-renormalizable 
terms introduce a dependence on the VEVs of the 
fields entering in the non-renormalizable couplings. Purely renormalizable 
Yukawa couplings are preferable, because the 
arbitrariness of such VEVs inevitably means that predictivity is lost. 
Furthermore, higher-order operators, 
such as those induced by the Fayet-Iliopoulos (FI)
breaking \cite{FayetIliopoulos}, 
are very model-dependent and introduce a high degree of uncertainty
in the computation. In addition, 
as emphasized in ref.~\cite{jhep},
their presence is not always allowed in string constructions. 
For example, in the 
$SU(3)_c\times SU(2)_L\times U(1)_Y$ model
of ref. \cite{Casas2}, those few 
non-renormalizable couplings that were allowed by gauge invariance 
were forbidden by the `stringy' selection rules. 

In specific orbifold constructions however, entirely renormalizable couplings
seem to be unable to explain the experimental data.
Summarizing the analyses in refs.~\cite{test,Faustino3},
for prime orbifolds the space group selection rule and the
need for a hierarchy forces the fermion mass matrices to be diagonal
at the renormalizable level. Thus in these cases 
the CKM parameters must arise at the non-renormalizable level. 
For non-prime orbifolds the mass matrices can actually be
non-diagonal at the renormalizable level,
however a reasonable CKM matrix again requires non-renormalizable
couplings.
In both cases, non-renormalizable couplings have to account for 
the masses of the first generation (coming  
from off-diagonal elements in the mass matrices),
but renormalizable couplings can still be responsible
for the masses of the second and third generation.
Under this assumption it was shown that, for a reasonable size and shape of the
compactified space, the $Z_3$, $Z_4$, $Z_6$-I, and possibly $Z_7$
orbifolds
can fit the physical quark and charged-lepton masses adequately, but that
the rest of the $Z_n$ orbifolds cannot. In all cases though, 
non-renormalizable terms were necessary to fit the physical masses.

How might we avoid having to use non-renormalizable terms?
One hint lies in the fact that their apparent necessity  
in the $Z_3$ orbifold
was deduced assuming a minimal $SU(3)_c\times SU(2)_L\times U(1)_Y$ 
scenario with a single generation of
supersymmetric Higgses ($H^u$, $H^d$).
This is the usual assumption in the context of the
Minimal Supersymmetric Standard Model (MSSM).
In addition, three Wilson lines were assumed.
This implies that the 27 twisted sectors of the
$Z_3$ orbifold are different, and then it is possible in principle
to assign a physical field to any fixed point. 
On the one hand, this is welcome since it
allows one to play
fully with suppression factors and hence to obtain a realistic 
fermion mass hierarchy, albeit with nonrenormalizable terms. 
On the other hand, it is problematic, since 
the existence of three families of 
$SU(3)_c\times SU(2)_L\times U(1)_Y$
is not guaranteed in all cases.
The latter is also true for all other orbifolds.

In contrast to the situation with three Wilson lines,
$Z_3$ orbifold models with two Wilson lines 
automatically have three families of everything, including Higgses. 
This is because, in addition to the overall factor of 3 coming from
the right-moving part of the untwisted matter, the twisted matter
comes in 9 sets with 3 equivalent sectors in each one, since there
are 27 fixed points. Consequently these compactifications 
are very interesting from a phenomenological
point of view. 
Indeed, several models with two Wilson lines, 
have been constructed with 
$SU(3)\times SU(2)\times U(1)^n$ 
observable gauge group and three families of particles \cite{Kim}.
In addition, using the FI 
mechanism \cite{FayetIliopoulos} it is possible to break
the original gauge group down to
$SU(3)_c\times SU(2)_L\times U(1)_Y$ \cite{Katehou,Casas2,Font,Casas3}.
Given these interesting properties,
the aim of the present paper is to analyze as systematically 
as possible the structure of Yukawa couplings in $Z_3$ orbifold models
with two Wilson lines, and to ascertain whether a purely renormalizable 
structure is possible.

To carry out the analysis, 
we first make the most natural assumption, 
which is that all three generations of supersymmetric
Higgses remain light ($H^u_i$, $H^d_i$), $i=1,2,3$\footnote{One might avoid 
the extra light Higgs generations if they became 
quite massive due to some kind of 
`asymmetric' breaking. However, this is certainly not natural in
these models. For example, one could generate a high mass for the
Higgses through trilinear couplings involving a field which
develops a VEV in order to cancel the FI D-term.
However, given the structure of Yukawa couplings in these models
(see Sect.~2), all three families of Higgses would become massive.}.
Indeed, in all the models that have been constructed, 
extra doublets are generically present at low energies.
This possibility also favours the unification
of gauge couplings in Heterotic Superstring constructions \cite{jhep}.
Importantly for our analysis, having three families of Higgses 
introduces more Yukawa couplings. 
The FI breaking provides a second important ingredient in our analysis
which appears at the renormalizable level.
Namely, after the gauge breaking some physical particles
appear combined with other states, and the
Yukawa couplings are modified in a well controlled way.
This, of course, introduces more flexibility in the
computation of the mass matrices.

All of these factors allow us to achieve our goal   
of obtaining realistic quark and lepton mass
matrices entirely at the {\it renormalizable} level,
and we will show below that the observed structure
in eqs.~(\ref{masses}) and (\ref{ckm}), can indeed be obtained.
Despite the modifications due to three Higgs families and 
FI mixing, the model retains a large degree of predictivity.
Indeed, if we assume one overall modulus we successfully 
predict two mixing angles and {\em four} masses 
(out of a total of three angles and nine quark and charged lepton 
masses).

$Z_3$ orbifold models with two Wilson lines
also allow us to attack the problem of 
neutrino masses and here we find some rather attractive features.
%
It turns out that we require a see-saw mechanism to generate small
enough neutrino masses. However, the Dirac mass matrices for 
neutrinos are already very small in these constructions because
the Yukawa couplings get the same kind of hierarchical suppression that 
appears in the 
quarks and charged leptons couplings. For the examples we present in the text, 
we will find that inserting the measured neutrino masses (assuming that
the mass-squared differences reflect the actual masses) then tells us 
that the see-saw scale has to be $\sim 10^4$ TeV. This generates an effective $\mu$-term 
of order 500 TeV which is only a couple of orders of magnitude above the weak scale.
We think that the {\em relative} 
closeness of the see-saw scale to the electroweak 
scale (compared to the usual see-saw mechanism) 
may be an important hint. First, it suggests that the three families of 
singlet responsible for the neutrino see-saw mechanism may get their 
VEVs when electroweak symmetry is broken. Second it suggests that 
the solution to the so called $\mu$-problem is that the Higgs fields couple to 
the same singlets as give a mass to the right handed neutrinos, possibly through 
slightly suppressed couplings.
The maximal mixing and rather mild hierarchies of the neutrino system can 
also be achieved with no other modifications. 

Of course, dangerous
flavour-changing neutral currents (FCNCs) may appear when
three generations
of Higgses are present \cite{Paschos}.
In general, the most stringent limit on flavour-changing processes comes
from the small value of the $K_L-K_S$ mass difference \cite{KK}.
So we need the extra Higgses to be
sufficiently  massive to suppress $\Delta S=2$ neutral currents 
contributions to agree with 
the experimental data \cite{Georgi,KK,Cheng}. 
The actual lower bound on Higgs masses depends on the particular texture
chosen for the Yukawa matrices, but can be as 
low as 120--200 GeV \cite{fcnc}.

The paper is organized as follows.
In Section~2 we will summarize first the structure of Yukawa couplings
between twisted fields in the $Z_3$ orbifold.
This will allow us to write the quark and lepton mass matrices.
Then, we will see that after the FI breaking,
these matrices are substantially modified.
In section 3 we analyze the quark masses and mixings and 
find that without the FI mixing, the CKM matrix cannot be reproduced, but that 
it can once we take account of the FI mixing. In section 4 we go on to consider the 
charged lepton and neutrino mass ratios and mixings. Section 5 deals with further 
constraints coming from the absolute values of the masses, and section 6 summarizes 
and collates the results.

\section{Yukawa couplings and mass matrices in the $Z_3$ orbifold}

As mentioned in the introduction, we are interested
in the case that the observable
matter lies entirely in the twisted sector, in order to get
a realistic mass hierarchy. Thus we first summarize the
characteristics of Yukawa couplings between twisted fields
in the $Z_3$ orbifold.

The $Z_3$ orbifold is constructed by dividing $R^6$ by the
$[SU(3)]^3$ root lattice
modded by the point group (P) with generator $\theta$,
where the action of $\theta$ on the lattice basis is
$\theta e_i=e_{i+1}$, 
$\theta e_{i+1}=-(e_i+e_{i+1})$ with $i=1,3,5$.
The two-dimensional sublattices associated to $[SU(3)]^3$ are shown
in Fig.~1.
Let us call $R_k=|e_k|$ and $\alpha_{kl}=\cos\theta_{kl}$,
where $e_ke_l=R_kR_l\cos\theta_{kl}$ and $k,l=1,...,6$.
The initial six-torus of the $Z_3$ orbifold 
has 21 degrees of freedom, however
taking into account the relations that
P-invariance imposes reduces these to only 9 as follows.
Indeed, preserving the magnitude of the lattice basis constrains $R_i=R_{i+1}$ and 
also $\alpha_{i,i+1}=-1/2$ or equivalently $\theta_{i,i+1}=2\pi/3$.
Preserving the angles under the action of P then enforces the two 
relations $\alpha_{i,j+1}+\alpha_{i+1,j}+\alpha_{i,j}=0$ and $\alpha_{i,j}=\alpha_{i+1,j+1}$.
Only the following nine deformation parameters are left \cite{test}:
\bea
R_1\ ,\,\ \ R_3\ ,\,\ \ R_5\ ,\,\ \ \alpha_{13}\ ,\,\ \ \alpha_{15}
\ ,\,\ \ \alpha_{35}\ ,\,\ \ \alpha_{14}\ ,\,\ \ \alpha_{16}
\ ,\,\ \ \alpha_{36}
\ .
\label{deformations}
\eea
In the $Z_3$ orbifold without deformations 
the angles between complex planes are vanishing, however this need not be the case.
The nine deformation parameters correspond
to the VEVs of nine singlet fields that appear in the 
spectrum of the untwisted sector (of the form $\psi^l_{-1/2}|0\rangle_L\times \tilde{\alpha}
^k_{-1}|0\rangle_R$) 
which have perturbatively flat potentials. These so-called 
moduli fields are usually denoted by $T$.

\begin{figure}
\centerline{
\epsfig{file=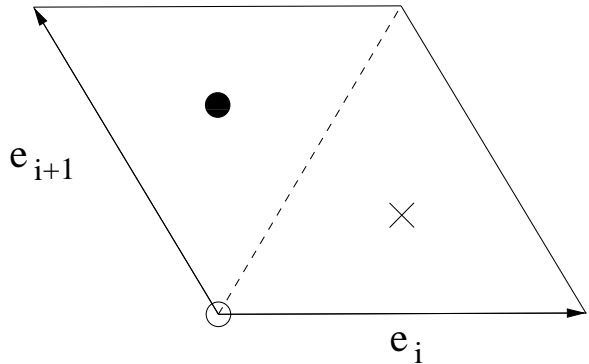,height=5cm,width=7.7cm,angle=0.2}}
\caption{Two dimensional sublattices ($i=1,3,5$) of the $Z_3$ orbifold.
The fixed point components are also shown.}
\label{fig1}
\end{figure}

In orbifold constructions, twisted strings appear 
attached to fixed points under the point group.
In the case of the $Z_3$ orbifold there are 27 fixed points under
P, and therefore there are 27 twisted sectors. 
We will denote the three fixed points of each two-dimensional
sublattice as shown in Fig.~1. 
For example, the (o\ x\ x) fixed point is in the position
$f_{oxx}=\frac{1}{3}(2e_3+e_4+2e_5+e_6)$.
It was shown in ref.~\cite{Dixon} that given two fields associated
to two fixed points $f_1$, $f_2$, they can only couple to
a unique third fixed point $f_3$ as a consequence of the so-called 
space group selection rules (thus there are 27$\times$ 27 = 729 
allowed Yukawa couplings). In particular, the
components of the three fixed points in each sublattice
must be either equal or different.
For example, if in one of the three sublattices the
components of the fixed points $f_1$ and $f_2$ are x, x, respectively,
the component of the third fixed point $f_3$ must also be x.
If the components for the first two fixed points are x, $\cdot$\ ,
then the component for the third fixed point must be o.

The expressions for the different Yukawa couplings can be found for
example in the Appendix of ref.~\cite{Faustino2}. They contain
suppressions factors that depend on the relative positions of the
fixed points to which the fields involved in the
coupling are attached (i.e. $f_1$, $f_2$, $f_3$) and on the
size and shape of the orbifold (i.e. the deformation parameters in 
eq.~(\ref{deformations})). In fact, it is possible to show that
only 14 couplings out of the 729 allowed are different. In the
particular of an orthogonal lattice, i.e. when the six angles in 
eq.~(\ref{deformations}) are zero, there are only 8 distinct couplings.
We will show in the following sections
that these three radii are sufficient to fit 
the whole quark and lepton masses, and so we henceforth 
restrict the discussion to this case. 

We begin by presenting the
general form of Yukawa coupling. They are given by the
Jacobi theta function,
\bea
Y_{\theta\theta\theta}=g\ N\sum_{{\bf u}\in Z^6} exp\ [-
2\pi
({\bf f}_{23}+{\bf u})^T M({\bf f}_{23}+{\bf u})]
\label{Yukawas}
\eea
with 
\bea
N={\sqrt V}\ \frac{3^{3/4}}{8\pi^3}\frac{\Gamma^6 (\frac{2}{3})}
{\Gamma^3 (\frac{1}{3})}
\label{normalization}
\eea
and
\bea
{M}=\left( \begin{array}{cccccc}
T_1 & -\frac{1}{2}T_1 & 0 & 0 & 0 & 0\\
-\frac{1}{2}T_1 & T_1 & 0 & 0 & 0 & 0\\
0 & 0 & T_3 & -\frac{1}{2}T_3 & 0 & 0\\
0 & 0 & -\frac{1}{2}T_3 & T_3 & 0 & 0\\
0 & 0 & 0 & 0 & T_5 & -\frac{1}{2}T_5\\
0 & 0 & 0 & 0 & -\frac{1}{2}T_5 & T_5\\
\end{array}\right) 
\ .
\label{matrix}
\eea
Here $g$ is the gauge coupling constant, 
$V$ is the volume of the unit cell for the $Z_3$ lattice,
and $T_i$ are the diagonal moduli whose real parts are associated
to the internal metric $g_{ii}=e_i e_i$,
$Re\ T_i=\frac{\sqrt 3}{16\pi^2}R_i^2$, and whose imaginary parts are
associated with the torsion. 
The fact that off-diagonal elements in the matrix $M$ are vanishing
is due to our assumption of an orthogonal lattice. 
The vector ${\bf f}_{23}$ represents the six
components of ($f_2-f_3$). The only eight inequivalent possibilities
are
\bea
(0,0,0,0,0,0)
\ ,
\label{zero}
\\
(\frac{1}{3},\frac{2}{3},0,0,0,0)\ ,\,\ \ 
(0,0,\frac{1}{3},\frac{2}{3},0,0)\ ,\,\ \ 
(0,0,0,0,\frac{1}{3},\frac{2}{3})
\ ,
\label{one}
\\
(\frac{1}{3},\frac{2}{3},\frac{1}{3},\frac{2}{3},0,0)\ ,\,\ \
(\frac{1}{3},\frac{2}{3},0,0,\frac{1}{3},\frac{2}{3})\ ,\,\ \
(0,0,\frac{1}{3},\frac{2}{3},\frac{1}{3},\frac{2}{3})\ ,
\label{two}
\\
(\frac{1}{3},\frac{2}{3},\frac{1}{3},\frac{2}{3},
\frac{1}{3},\frac{2}{3})
\ ,
\label{three}
\eea
corresponding to three fields with the same
fixed point components in three sublattices, two sublattices, one
sublattice, and no sublattice, respectively.

It turns out that for reasonable values of the moduli, 
$T_i={\cal O} (1)$,
all terms in the sum of eq.~(\ref{Yukawas})
are negligible with respect to those corresponding
to the shortest distance between fixed points.
For example, for the third vector in eq.~(\ref{one})
the sum, say $\varepsilon_5$, is given by 
%
\bea
\varepsilon_5=3\ e^{-\frac{2\pi}{3}T_5}\
(1+ 6\ e^{-{2\pi}T_1}+ 6\ e^{-{2\pi}T_3}+\dots\ )
\ ,
\label{epsilon1}
\eea
where the dots denote terms with larger suppression factors, which
can therefore be approximated as 
$\epsilon_5\approx 3\ e^{-\frac{2\pi}{3}T_5}$. 
In the next section we will obtain $T_5\sim 1.95$ and therefore
$\epsilon_5\sim 0.05$. 
Taking this into account, the sums in eq.~(\ref{Yukawas})
corresponding to eqs.~(\ref{zero}-\ref{three}) can be
approximated respectively as
%
\bea
1
\ ,
\label{epsilonzero}
\\
\varepsilon_1\ ,\,\ \
\varepsilon_3\ ,\,\ \
\varepsilon_5
\ ,
\label{epsilonone}
\\
\varepsilon_{13}
\ ,\,\ \
\varepsilon_{15}
\ ,\,\ \
\varepsilon_{35}
\ ,
\label{epsilontwo}
\\
\varepsilon_{135}
\ ,
\label{epsilonthree}
\eea
where $\varepsilon_i=3\ e^{-\frac{2\pi}{3}T_i}$,
$\varepsilon_{ij}=\varepsilon_i \varepsilon_j$,
and 
$\varepsilon_{135}=\varepsilon_1\varepsilon_3 \varepsilon_5$.
Finally we remark that in the case without deformations,
i.e. assuming an orthogonal lattice with $T_i=T$, there are only
four different couplings, with   
the sums in eq.~(\ref{Yukawas}) given by
1, $\varepsilon$, $\varepsilon^2$, $\varepsilon^3$, where
$\varepsilon=3\ e^{-\frac{2\pi}{3}T}$.
Although we mentioned above that three radii are sufficient to fit
the quark and lepton masses, in actual fact this 
is possible with only one degenerate radius and no deformations 
at all, as we shall see in the following section. 
This is of course the most predictive assumption. Conversely, 
if we begin by allowing three different radii, once we fit the 
quark and lepton masses, we determine them to be very close.

\subsection{Mass matrices before the Fayet-Iliopoulos breaking}

With these results we can now turn to the analysis of mass matrices in
constructions with two Wilson lines. In what follows we shall consider 
a particular assignment of Standard Model
particles to different fixed points
and examine the predictions from the 
subsequent Yukawa 
couplings\footnote{Note that generally the Standard Model particles {\em do} 
belong to different fixed points with such issues as 
anomaly cancellation being fixed automatically by the string construction.}.  
This is 
a `bottom up' approach in that we do not explicitly construct the 
models but are asking which assignment is appropriate for the observed masses and 
mixings.

Let us first study the situation 
before taking into account the effect of the FI breaking.
Let us suppose that the two non-vanishing Wilson lines
($a_1$, $a_3$) correspond to the first and second sublattices.
Then the 27 twisted sectors come in nine sets with three equivalent
sectors in each one.
The three generations of matter (including Higgses) 
correspond to changing the third sublattice component
(x, $\cdot$\ , o) of the fixed point whilst keeping the other two fixed.
Consider for example the following assignments of observable matter to
fixed point components in the first two sublattices;
\bea
Q\ \,\,\,\, o\ o\ \,\,\,\,\,\,\,\,\,\,\,\, \ u^c\ \,\,\,\, o\ o
\ \,\,\,\,\,\,\,\,\,\,\,\, \ 
d^c\ \,\,\,\, x\ o
\nonumber
\\
L\ \,\,\,\, {\cdot}\ {\cdot}
\ \,\,\,\,\,\,\,\,\,\,\,\, \
e^c\ \,\,\,\, {\cdot}\ x
\ \,\,\,\,\,\,\,\,\,\,\,\, \
\nu^c\ \,\,\,\, x\ x
\nonumber
\\
H^u\ \,\,\,\, o\ o
\ \,\,\,\,\,\,\,\,\,\,\,\, \
H^d\ \,\,\,\, {\cdot}\ o
\ \,\,\,\,\,\,\,\,\,\,\,\, 
\label{assignments}
\eea
In this case the up- and down-quark mass matrices, assuming three 
different radii, are given by
%
\bea
M^u=gNA^u\ \,\,\,\,\ ,\ \,\,\,\, 
M^d=gN\varepsilon_1 A^d
\ ,
\label{quarkmasses}
\eea
where  
\bea
A^{u}=\left( \begin{array}{ccc}
v_1^{u} & v_3^{u}\varepsilon_5  & v_2^{u}\varepsilon_5 \\
v_3^{u}\varepsilon_5  & v_2^{u} & v_1^{u}\varepsilon_5 \\
v_2^{u}\varepsilon_5  & v_1^{u}\varepsilon_5  & v_3^{u}
\end{array}\right)
\,\,\,\,\ , \,\,\,\,
A^{d}=\left( \begin{array}{ccc}
v_1^{d} & v_3^{d}\varepsilon_5  & v_2^{d}\varepsilon_5 \\
v_3^{d}\varepsilon_5  & v_2^{d} & v_1^{d}\varepsilon_5 \\
v_2^{d}\varepsilon_5  & v_1^{d}\varepsilon_5  & v_3^{d}
\end{array}\right)
\ .
\label{quarkmasses2}
\eea
Here
$v^u_i$, $v^d_i$ denote the VEVs of the Higgses 
$H^u_i$, $H^d_i$ respectively.
For simplicity we will assume for the moment that these VEVs,
as well as those of the moduli $T_i$, are real. 
Of course, in general they can be complex numbers, and
later on we will address the importance that this may have for
CP violation.

The elements in the above matrices can be obtained straightforwardly.
For example, if the Higgs $H^u_1$ corresponds to 
(o,o,o), then since the three generations of (3,2) quarks $Q$ 
correspond to (o,o,(o,x,$\cdot$)) and the three generations
of ($\bar 3$,1) quarks $u^c$ to (o,o,(o,x,$\cdot$)),
there are only three allowed couplings,
\begin{eqnarray}
&&\mbox{(o,o,o)(o,o,o)(o,o,o)}\ ,\nonumber \\
&&\mbox{(o,o,o)(o,o,x)(o,o,$\cdot$)}\ , \nonumber \\
&&\mbox{(o,o,o)(o,o,$\cdot$)(o,o,x)}\ . \nonumber
\end{eqnarray}
\noindent The corresponding suppression factors are given by 1,
$\varepsilon_5$, $\varepsilon_5$ respectively,
and are associated with the elements 11, 23, 32 in the
matrix $M^u$. Using the same argument, we obtain the neutrino mass
matrix $M^{\nu}$, as well as the charged-lepton mass matrix  $M^{e}$,
\bea
M^{\nu}=\varepsilon_{1}\varepsilon_{3}\ M^u
\ ,
\,\,\
M^{e}=\frac{\varepsilon_{3}}{\varepsilon_1}\ M^d
\ .
\label{leptonmasses}
\eea

\subsection{Mass matrices after the Fayet-Iliopoulos breaking}

Let us now study how the previous results are modified when
one takes into account the FI breaking.
As mentioned in the introduction, some scalars,
in particular $SU(3)\times SU(2)$ singlets, develop large VEVs,
of the order of $10^{16-17}$ GeV, in order
to cancel the FI D-term generated by the anomalous $U(1)$. 
The VEVs of these fields, which 
we shall denote $C_i$,
break the original gauge group
$SU(3)\times SU(2)\times U(1)^n$ 
down to $SU(3)_c\times SU(2)_L\times U(1)_Y$.

After the breaking, 
many particles, which we will refer to as $\xi$, are expected to 
acquire a high mass because
of the generation of effective mass terms. These come for example
from operators of the type $C\xi\xi$.
In this way vector-like triplets and doublets and also singlets become 
heavy and disappear from the low-energy spectrum. 
This is the type of extra matter that typically 
appears in orbifold constructions. 
The remarkable point is that the Standard Model matter remain massless,
surviving through certain combinations with other 
states \cite{Kim}-\cite{Casas3}.
Let us consider the simplest example, a model with the
Yukawa couplings
\bea
C_1 \xi_1 f
\ ,
\,\,\
C_2 \xi_1 \xi_2 \, , 
\label{massive}
\eea
where $f$ denotes a Standard Model field,
$\xi_{1,2}$ denote two extra matter fields (triplets, doublets or singlets),
and $C_{1,2}$ are the fields developing large VEVs
denoted by $\langle C_{1,2} \rangle = c_{1,2}$.
It is worth noting here that $f$ can be an $u^c$, $d^c$, $L$, 
$\nu^c$ or $e^c$
field, but not a $Q$ field. This is because in these orbifold models no 
extra (3,2) representations are present, and therefore the
Standard Model field $Q$ cannot mix with other representations through
Yukawas.

Clearly the `old' physical particle $f$ will combine with
$\xi_{1,2}$. It is now straightforward to diagonalize
the mass matrix arising from the mass terms in eq.~(\ref{massive}) to
find two very massive and one massless combination. 
The latter is given by
\bea
f'\equiv \frac{1}{\sqrt{|c_{1}|^{2}+|c_{2}|^{2}}}
\left(c_2^* f - c_1^* \xi_2\right)
\ .
\label{massless}
\eea
Notice for example that the mass terms (\ref{massive})
can be rewritten as
$\sqrt{|c_{1}|^{2}+|c_{2}|^{2}}\ \xi_1\xi'_2$,
where $\xi'_2\equiv
\frac{1}{\sqrt{|c_{1}|^{2}+|c_{2}|^{2}}}
\left(c_1 f + c_2 \xi_2\right)$.
Indeed the unitary combination is the massless field in eq.~(\ref{massless}).
The Yukawa couplings and hence mass matrices of the effective low energy 
theory are modified accordingly. 
For example, consider a model where we begin with a 
Yukawa coupling 
$H Q f$.
Since we have
\bea
f=\frac{1}{\sqrt{|c_{1}|^{2}+|c_{2}|^{2}}}
\left(c_2 f' + c_1^* \xi'_2\right)
\ ,
\label{oldfield}
\eea
then the `new' coupling (involving the light state) will 
be\footnote{We should add that the coupling
$H Q \xi_2$, which would induce another contribution to 
$H Q f'$, is not in fact allowed. For this to be the case
the fields $\xi_2$ and $f$ would have had 
to have exactly the same $U(1)^n$ charges.
This is not possible since different particles all have different
gauge quantum numbers.}
\[\frac{c_2}{\sqrt{|c_{1}|^{2}+|c_{2}|^{2}}} H Q f' \, .\]

The situation in realistic models is more involved since
the fields appear in three copies. Thus the mass matrix
for the example in eq.(\ref{massive}) is (using the results above)
\bea
g N\ 
(\xi_1^1\ \xi_1^2\ \xi_1^3)\left\{ 
\varepsilon'
\left(\begin{array}{ccc}
c_{1}^1 & c_{1}^3\varepsilon_{5}  & c_1^{2}\varepsilon_{5} \\
c_{1}^3\varepsilon_5  & c_{1}^2 & c_{1}^1\varepsilon_5 \\
c_{1}^2\varepsilon_5  & c_{1}^1\varepsilon_5  & c_{1}^3
\end{array}\right) 
\left(\begin{array}{c}
f^{1} \\
f^{2} \\
f^{3} 
\end{array}\right) 
\ +\ 
\varepsilon''
\left( \begin{array}{ccc}
c_{2}^1 & c_{2}^3\varepsilon_{5}  & c_2^{2}\varepsilon_{5} \\
c_{2}^3\varepsilon_5  & c_{2}^2 & c_{2}^1\varepsilon_5 \\
c_{2}^2\varepsilon_5  & c_{2}^1\varepsilon_5  & c_{2}^3
\end{array}\right)
\left(\begin{array}{c}
\xi_{2}^1 \\
\xi_{2}^2 \\
\xi_{2}^3  
\end{array}\right) 
\right\}
\ ,
\label{mixings}
\eea
where $\varepsilon', \varepsilon''$ can take different values
\bea
\varepsilon' , \varepsilon'' \equiv
1\ ,\ \varepsilon_1\ ,\  \varepsilon_3\ ,\ \varepsilon_1 \varepsilon_3\ \ ,
\label{epsilons}
\eea
depending on the particular case.
For example, if the field $f$ corresponds to the down quark with
the assignment as in eq.~(\ref{assignments}),
and the fields $C_1, C_2, \xi_1, \xi_2$ have the assignments for the
first two sublattices
(x o), (o x), (x o), ($\cdot$ $\cdot$), respectively,
then $\varepsilon'=1$ and $\varepsilon''=\varepsilon_1 \varepsilon_3$.

Now, in order to simplify the 
analysis, let us consider
the following VEVs for the $C_{1,2}^i$ fields\footnote{In principle 
we are allowed to do this since the cancellation of the
FI D-term only imposes $\sum_i Q_i^{(a)} (|c_i^1|^2 + 
|c_i^2|^2 + |c_i^3|^2)= const$,
where $Q_i^{(a)}$ are the charges of the $C_i$ fields
under the anomalous $U(1)$, and therefore flat directions arise.
As for the $T$-moduli, 
these VEVs can eventually be determined dynamically through 
supersymmetry breaking. 
For attempts in this direction see e.g. ref.~\cite{Giedt2}.};
\bea
c_1^1\equiv c_1\ , \,\,\,\,\,\,\,\,\,\,\,\,\,\,\,\, 
c_1^2=c_1^3=0
\ ,
\nonumber
\\
c_2^1=c_2^3=0\ , \,\,\,\,\,\,\,\,\,\,\,\,\,\,\,\,
c_2^2\equiv c_2
\ .
\label{simplify}
\eea
Then eq.~(\ref{mixings}) gives rise to the mass terms
\bea
\sqrt{|\hat c_{1}|^{2}+ |\hat c_{2}\varepsilon_5|^{2}}\ \xi_1^1 \xi_2'^{1}
\ +\
\sqrt{|\hat c_{1}\varepsilon_5|^{2}+|\hat c_{2}|^{2}}\ \xi_1^2 \xi_2'^2
\ +\
\varepsilon_5
\sqrt{|\hat c_{1}|^{2}+|\hat c_{2}|^{2}}\ \xi_1^3 \xi_2'^3
\ ,
\label{mixings2}
\eea
where 
\bea
\xi_2'^1 & \equiv &
\frac{1}{\sqrt{|\hat c_{1}|^{2}+|\hat c_{2}\varepsilon_5|^{2}}}
\left(\hat c_1 f_1 + \hat c_2 \varepsilon_5 \xi_2^3\right)
\ ,
\nonumber\\ 
\xi_2'^2 & \equiv &
\frac{1}{\sqrt{|\hat c_{1}\varepsilon_5|^{2}+|\hat c_{2}|^{2}}}
\left(\hat c_1 \varepsilon_5 f_3 + \hat c_2 \xi_2^2\right)
\ ,
\nonumber
\\ 
\xi_2'^3 & \equiv &
\frac{1}{\sqrt{|\hat c_{1}|^{2}+|\hat c_{2}|^{2}}}
\left(\hat c_1 f_2 + \hat c_2 \xi_2^1\right)
\ ,
\label{mixings3}
\eea
and 
\bea
\hat c_1\equiv \varepsilon' c_1 \,\,\, \ , \,\,\,\,\,\,\,\,\,
\hat c_2\equiv \varepsilon'' c_2
\ .
\label{mixings4}
\eea
Following the discussion for
eqs.~(\ref{massless}) and (\ref{oldfield})
we can deduce straightforwardly
that the new mass matrices for the quarks are 
\bea
{\cal M}^u=a^{u^c} M^u  B^{u^c}
\ \,\,\,\,\ ,\ \,\,\,\, 
{\cal M}^d=a^{d^c} M^d  B^{d^c}
\ ,
\label{newquarkmasses}
\eea
where\footnote{Note that, although $c_{1,2}$ are in general
complex VEVs, they only introduce a global and therefore unphysical 
phase in the mass matrix. Of course, this is an artifact of the direction
(\ref{simplify})
choosen to cancel the FI D-term. More complicated directions
would give rise in principle to a contribution to the
CP phase.
This mechanism to generate the CP phase through the VEVs of the
fields cancelling the FI D-term was used first, in the context of
non-renormalizable couplings, in ref.~\cite{test}.
For a recent analysis see ref.~\cite{Giedt}.} 
\bea
a^f=
\frac{\hat c_{2}^f}{\sqrt{
|\hat c_{1}^f|^{2}+
|\hat c_{2}^f|^{2}
}}
\ ,
\label{af}
\eea
$M^{u}, M^d$ are given in eq.~(\ref{quarkmasses}),
and
%
\bea
B^f=
\left( \begin{array}{ccc}
\beta^f \varepsilon_5   & 0 & 0\\
0 & 1 & 0\\
0 & 0 & {\alpha^f}/{\varepsilon_5}
\end{array}\right)
\ ,
\label{matrixB}
\eea
with
\bea
\alpha^f =\varepsilon_5\ \sqrt{\frac{|\hat c_{1}^f|^{2}+
|\hat c_{2}^f|^{2}}{|\hat c_{1}^f\varepsilon_5|^{2}+
|\hat c_{2}^f|^{2}}}\ , \,\,\,\,\,\, 
\beta^f =
\sqrt{\frac{|\hat c_{1}^f|^{2}+
|\hat c_{2}^f|^{2}}{|\hat c_{1}^f|^{2}+
|\hat c_{2}^f\varepsilon_5|^{2}}}
\ .
\label{alpha}
\eea
Here 
we have already taken into account
that different fields will couple to  
different $C_i$ fields, and therefore we will generally 
have 
$c_{1,2}^{u^c}\neq c_{1,2}^{d^c}$.

Concerning the leptons, we have two possible structures
for the mass matrices. If we assume that only the fields
$\nu^c$ and $e^c$ mix with other representations through Yukawas, the
situation is similar to that above, generating the following
mass matrices;
\bea
{\cal M}^{\nu}=a^{{\nu}^c} M^{\nu}  B^{\nu^c}
\ \,\,\,\,\ ,\ \,\,\,\, 
{\cal M}^e=a^{e^c} M^e  B^{e^c}
\ ,
\label{newleptonmasses}
\eea
where $M^{\nu}, M^e$ are given in eq.~(\ref{leptonmasses}).
However, in principle, the left handed leptons $L$
may also mix with other representations, which instead gives rise
to the matrices
\bea
{\cal M}^{\nu}=a^{L} a^{{\nu}^c} 
B^L 
M^{\nu}  B^{{\nu}^c}\,\,\,
\ ,\,\,\,
{\cal M}^e=a^{L} a^{e^c} B^{L} 
M^e  B^{e^c}
\ .
\label{newleptonmasses2}
\eea
We will find that the second possibility is the one which is consistent with 
the observed masses.

For a given field
there are basically three patterns for the values of $\alpha$
and  $\beta$. When 
$\hat c_{1} \sim \hat c_{2}$ one obtains
$\alpha\sim \varepsilon_5$ and $\beta\sim 1$, since 
in both denominators in eq.~(\ref{alpha}) 
the term with $\varepsilon_5$ is negligible.
In addition $a\sim 1$.
The possibility with 
$\hat c_{1} \ll \hat c_{2}$ may also be present and turns out
to give a trivial result.
For example, this is the case when $\varepsilon''=1$,
$\varepsilon'=\varepsilon_1\varepsilon_3$, and therefore using
eq.(\ref{mixings4}) one obtains
$\hat c_{2}=c_2$ and $\hat c_{1}=c_1\varepsilon_1\varepsilon_3$.
Since one expects $c_1\sim c_2$, as obtained in 
explicit models \cite{Kim}-\cite{Casas3},
$\hat c_{1}$ is much smaller than $\hat c_{2}$.
As a consequence, $\alpha\approx \varepsilon_5$, $\beta\approx 
1/\varepsilon_5$, and therefore $B\sim 1$.
Finally, the third pattern arises when
$\hat c_{1} \gg \hat c_{2}$, i.e. $\varepsilon' \gg \varepsilon''$.
In this case one gets
\bea
\alpha\sim 1
\ \,\,\,\, , \,\,\,\,\,\,\,\,\,\, \beta\sim 1
\ .
\label{alphabeta}
\eea
In addition,
\bea
a\sim \frac{c_2}{c_1}\frac{\varepsilon''}{\varepsilon'}
\sim \frac{\varepsilon''}{\varepsilon'}
\ ,
\label{af2}
\eea
%
where taking into account eq.~(\ref{epsilons}),
the above ratio 
can take the values
$\varepsilon''/
\varepsilon'=\varepsilon_1, \varepsilon_3, \varepsilon_1\varepsilon_3$.
An example of the above situation 
is given by
$\varepsilon'=1$ and 
$\varepsilon''=\varepsilon_1\varepsilon_3$.
There is a subtlety in some cases, as for example when
$\varepsilon'=1$ and 
$\varepsilon''=\varepsilon_1$,
since then
\bea
\alpha \sim \varepsilon_5\ \sqrt{\frac{|c_{1}|^{2}}
{|c_{1}\varepsilon_5|^{2}+
|c_{2}\varepsilon_1|^{2}}}
\sim \frac{\varepsilon_5}{\sqrt{|\varepsilon_5|^{2}+|\varepsilon_1|^{2}}}
\ .
\label{alpha2}
\eea
Thus $\alpha\sim \varepsilon_5/\varepsilon_1$
if $\varepsilon_1 >> \varepsilon_5$, and therefore 
the element 33 in $B$ above would be 
$1/\varepsilon_1$ instead of $1/\varepsilon_5$.
On the other hand, one does not expect an `asymmetric' supersymmetry 
breaking to occur naturally, and therefore
the moduli $T_i$ and hence the $\varepsilon_i$ 
should be of the same order.
As we have already mentioned, we will determine 
the $\varepsilon_i$ in the next section
using the available information from quark masses and mixing angles,
eqs.~(\ref{masses}) and (\ref{ckm}), and
will indeed find that the most attractive solution is 
where all of them are of the same order.
Finally, it is worth noting that the pattern
giving rise to values (\ref{alphabeta})
will be the most successful one for the 
quark masses, as we will discuss in the next section. 
For the leptons the first pattern will also be interesting,
as we will see in section~4.

Let us add that permutations in the diagonal elements of
the above matrix (\ref{matrixB})
are also possible.
This is because as well as the direction in 
eq.(\ref{simplify}) we have five additional non-trivial simple possibilities.
For example, we could instead have 
assumed $c_1^1=c_1$, $c_2^3=c_2$, with all other VEVs
vanishing, and then 
the elements 22 and 33 in (\ref{matrixB}) would have been permuted.
We will see below that this additional degree of freedom
is helpful in reproducing the observed neutrino masses. 

\section{Quark mass ratios and mixings}

One nice feature of the models we are interested in is that the Yukawas
have a similar form modulo overall factors. This means that, for example,
the mass ratios of the first/second generation downs are related to
those of the first/second generation leptons, and there is considerable
predictive power. Therefore it makes sense to determine mass ratios
and mixings before dealing with the various overall prefactors. 

We will try first to extract information from eq.~(\ref{quarkmasses}).
We will show that although the observed quark mass ratios and
Cabibbo angle can be reproduced correctly, the 13 and 23 elements
of the CKM matrix in eq.~(\ref{ckm}) cannot be obtained.
Fortunately, this is not the end of the story.
We will see that the matrices obtained after FI breaking, 
modified with the contribution in eq.~(\ref{matrixB}) will improve the
result, giving rise to the correct elements for the CKM matrix.

\subsection{Before the Fayet-Iliopoulos breaking}

First consider the quark mass matrices (\ref{quarkmasses})
before taking into account the impact of the FI breaking.
Clearly these matrices are very
constrained, and indeed it is possible to see immediately that they
are incompatible with a successful CKM matrix.

In order to prove this we can use the following procedure to find the
CKM matrix.
The symmetric matrix $A$ of eq.~(\ref{quarkmasses2})
can be diagonalized by a matrix $V$
%
\bea
{A}_{diag}=V A\ V^{T}
\ ,
\label{diagonal}
\eea
where we can define $V$
as orthogonal rotations,
\bea
V=R_{12}\ R_{23}\ R_{13}\ ,
\label{orthogonal}
\eea
through angles \( \phi _{12},\, \phi _{23},\, \phi _{13}. \) Now
we can write the rotations as an expansion in \( \varepsilon_5 \),
\bea
\sin \phi _{ij}=a_{ij}\varepsilon_5 +b_{ij}\varepsilon_5^{2}+
c_{ij}\varepsilon_5^{3}+\ldots \ ,
\label{expansion}
\eea
and solve for $(V {A} V^T)_{ab} =0$, where $a\neq b$, 
order by order in $\varepsilon_5$ to the
desired accuracy, deducing the coefficients \( a_{ij}, \) \( b_{ij} \),
\( c_{ij} \) as we go along. The CKM matrix is then 
\bea
V_{CKM}
= V_{U}\ V_{D}^{T}
\ .
\label{ckmmatrix}
\eea

Assuming \( \varepsilon  \) is sufficiently small, which we check
presently, to first order in epsilon the eigenvalues and diagonalization
of \( A \) goes as 
\bea
V A\ V^{T}=\left( \begin{array}{ccc}
v_{1} & 0 & 0\\
0 & v_{2} & 0\\
0 & 0 & v_{3}
\end{array}\right) 
+ {\cal O}(\varepsilon_5 )
\ ,
\label{another}
\eea
where
\bea
V=\left( \begin{array}{ccc}
1 & \frac{ v_{3}}{v_{1}-v_{2}}\varepsilon_5 & 
\frac{v_{2}}{v_{1}-v_{3}}\varepsilon_5\\
-\frac{ v_{3}}{v_{1}-v_{2}}\varepsilon_5 & 1 & 
\frac{v_{1}}{v_{2}-v_{3}}\varepsilon_5\\
-\frac{v_{2}}{v_{1}-v_{3}}\varepsilon_5 & 
-\frac{ v_{1}}{v_{2}-v_{3}}\varepsilon_5 & 1
\end{array}\right) 
+ {\cal O}(\varepsilon_5 ^{2})
\ .
\label{another2}
\eea
Without FI breaking therefore, the mass hierarchies must be provided
entirely by the Higgs VEVs, and we have 
\bea
\{v_{1}^{u},\, v_{2}^{u},\, v_{3}^{u}\} & 
\propto  & \{m_{u},\, m_{c},\, m_{t}\}
\ ,
\nonumber\\
\{v_{1}^{d},\, v_{2}^{d},\, v_{3}^{d}\} & 
\propto  & \{m_{d},\, m_{s},\, m_{b}\}
\label{proportional}
\ .
\eea
%
The CKM matrix is then given by 
\bea
V_{CKM}
\approx \left( \begin{array}{ccc}
1 & \varepsilon_5 \left( \frac{m_{b}}{m_{s}}
-\frac{m_{t}}{m_{c}}\right) & 
\varepsilon_5 \left( \frac{m_{s}}{m_{b}}-\frac{m_{c}}{m_{t}}\right) \\
-\varepsilon_5 \left( \frac{m_{b}}{m_{s}}-\frac{m_{t}}{m_{c}}\right)  
& 1 & \varepsilon_5 \left( \frac{m_{d}}{m_{b}}-\frac{m_{u}}{m_{t}}\right) \\
-\varepsilon_5 \left( \frac{m_{s}}{m_{b}}-\frac{m_{c}}{m_{t}}\right)  
& -\varepsilon_5 \left( \frac{m_{d}}{m_{b}}-\frac{m_{u}}{m_{t}}\right)  & 1
\end{array}\right) +O(\varepsilon_5 ^{2})
\ .
\label{another3}
\eea
From \( \left( V_{CKM}\right) _{12}\approx 0.22 \), and taking 
for example $m_{u}=4.5$ MeV,
$m_{d}=8.5$ MeV, $m_{s}=100$ MeV, $m_{c}=1.35$GeV, $m_{b}=4.5$ GeV
and $m_{t}=175$ GeV we determine 
\( \varepsilon_5 \approx 2.6\times 10^{-3} \)
(warranting our assumption of small \( \varepsilon  \)) and hence
\bea
\left( V_{CKM}\right) _{13} \approx   3.8\times 10^{-5}\ \,\, ,
\,\,\,
\left( V_{CKM}\right) _{23}  \approx   4.8\times 10^{-6}
\ .
\label{morenumbers}
\eea
Thus, quite generally, the model without FI breaking fails already
at the quark mixing stage.

\subsection{After the Fayet-Iliopoulos breaking}

As discussed in Subsection~2.2, in all realistic models 
constructed 
the standard-model matter survives through certain
combinations with other states \cite{Kim}-\cite{Casas3}.
Taking this into account we will see that the above results
can be modified. In particular, 
it will be possible to get the right spectrum and a CKM matrix with
the right form.
For the quarks the final mass matrices are now given as in 
eq.~(\ref{newquarkmasses})
%
\bea
{\cal M}^u=gNa^{u^c} A^u  B^{u^c}
\
\,\, ,\ \,\, 
{\cal M}^d=gN\varepsilon_1a^{d^c} A^d  B^{d^c}
\
\ ,
\label{finalud}
\eea
where 
\bea
A\  B
=
\left( \begin{array}{ccc}
v_1 \varepsilon_5 \beta & v_3\varepsilon_5  & v_2\alpha \\
v_3\varepsilon_5^2 \beta  & v_2 & v_1\alpha \\
v_2\varepsilon_5^2 \beta  & v_1\varepsilon_5  & 
v_3\alpha/\varepsilon_5
\end{array}\right)
\ ,
\label{define}
\eea

As discussed below eq.~(\ref{matrixB}), 
there are three possible patterns for $B$ depending
on the particular Yukawa couplings producing the combination of the
fields. The matrix 
$B$ with values as in eq.~(\ref{alphabeta}) will be the interesting one. 
The reason this choice works well is that
the hierarchy in the masses is now driven by the $B$ matrix
which aligns the Yukawas into hierarchical columns.
%
To leading order in \( \varepsilon_5  \) the eigenvalues 
of the matrix (\ref{define})
are given
by 
\bea
\left\{v_{1}\varepsilon_5 \beta
-\frac{(v_{3} \alpha)^{2} \varepsilon ^{3}_5 }{v_{2}}\ ,
\, v_{2}\ ,\, \frac{v_{3} \alpha }{\varepsilon_5 }\right\}
\ .
\label{eigenvaluesud}
\eea
As we will need a 
slight hierarchy in the VEVs, $v_3 > v_2 > v_1$, 
we have kept the \( \varepsilon_5^{3} \) term in the lightest eigenvalue.
We will see that this will be negligible for the down quark 
but not for the up quark.
With the approximate eigenvalues above we then have
\bea
\left\{v_{1}^u\ , v_{2}^u\ , v_{3}^u\right\}
\propto
\left\{\frac{1}{\varepsilon_5 \beta^{u^c}} 
\left(m_u+ \varepsilon_5^2\ \frac{m_t^2}{m_c}\right), 
m_c, \frac{m_t \varepsilon_5}{\alpha^{u^c}}\right\}
\ ,
\label{eigenvaluesudapprox1}
\\
%
\left\{v_{1}^d\ , v_{2}^d\ , v_{3}^d\right\}
\propto
\left\{\frac{1}{\varepsilon_5 \beta^{d^c}} 
\left(m_d+ \varepsilon_5^2\ \frac{m_b^2}{m_s}\right), 
m_s\ , \frac{m_b \varepsilon_5}{\alpha^{d^c}}\right\}
%
\ .
\label{eigenvaluesudapprox2}
\eea


In order to find the CKM matrix we can use the same procedure as 
in the previous subsection.
The full expression for the CKM matrix is independent of \( \alpha's  \)
and \( \beta's  \), and
in fact the role of the $B$ matrices is simply to modify the eigenvalues.
To leading order in \( \varepsilon_5  \) the CKM matrix becomes
\bea
V_{CKM}
=\left( \begin{array}{ccc}
1 & \left( \frac{v_{3}^{d}}{v_{2}^{d}}-
\frac{v_{3}^{u}}{v_{2}^{u}}\right) 
\varepsilon_5  & \left( \frac{v_{2}^{d}}{v_{3}^{d}}
-\frac{v_{2}^{u}}{v_{3}^{u}}\right) \varepsilon_5 \\
- \left( \frac{v_{3}^{d}}{v_{2}^{d}}-
\frac{v_{3}^{u}}{v_{2}^{u}}\right) 
\varepsilon_5  & 1 & \left( \frac{v_{1}^{d}}{v_{3}^{d}}-
\frac{v_{1}^{u}}{v_{3}^{u}}\right) \varepsilon_5 \\
-\left( \frac{v_{2}^{d}}{v_{3}^{d}}-
\frac{v_{2}^{u}}{v_{3}^{u}}\right) \varepsilon_5
& -\left( \frac{v_{1}^{d}}{v_{3}^{d}}-
\frac{v_{1}^{u}}{v_{3}^{u}}\right) \varepsilon_5  & 1
\end{array}\right) + {\cal O}(\varepsilon_5 ^{2}) 
\ ,
\label{ckmfinal}
\eea
%
For the calculation of 
the 13 element of the CKM matrix, $(V_{CKM})_{13}$,
 we require an accuracy of
\( \varepsilon_5 ^{2} \). The additional piece is  given by 
\bea
(V_{CKM})_{13}=\left( 
\frac{v_{2}^{d}}{v_{3}^{d}}-
\frac{v_{2}^{u}}{v_{3}^{u}}\right) 
\varepsilon_5 +\left( \frac{v_{1}^{u}}{v_{2}^{u}}-
\frac{v_{1}^{d}}{v_{3}^{d}}\frac{v_{3}^{u}}{v_{2}^{u}}\right) 
\varepsilon_5 ^{2}+\ldots
\ .
\label{accuracy}
\eea
Using eqs.~(\ref{eigenvaluesudapprox1}) and (\ref{eigenvaluesudapprox2})
we obtain 
%
\bea
(V_{CKM})_{12}  &=&  -\varepsilon_5 ^{2}
\left( \frac{m_{t}}{
m_{c}}\ \frac{1}{\alpha ^{u^c}}-\frac{m_{b}}{
m_{s}}\ 
\frac{1}{\alpha ^{d^c}}\right) \ , 
\nonumber\\
(V_{CKM})_{23}
 &=&  \frac{1}{\varepsilon_5 }
\left( \frac{1}{m_b}
\left[
{m_{d}}
+
\varepsilon_5^5\ \frac{m_b^2}{m_s}
\right]
\ 
\frac{\alpha ^{d^c}}{\beta ^{d^c}}
-\frac{1}{m_t}\left[
{m_{u}}
+
\varepsilon_5^5\ \frac{m_t^2}{m_c}
\right]
\ \frac{\alpha ^{u^c}}{\beta ^{u^c}}
\right) 
\ ,
\nonumber\\
(V_{CKM})_{13} & =&  \left(\frac{m_{s}}{m_{b}}\ \alpha ^{d^c}-
\frac{
m_{c}}{m_{t}}\ \alpha ^{u^c}\right)
\nonumber\\
&& -\ 
\varepsilon_5 \left(
\frac{1}{m_{b}}\
\left[m_d+ \varepsilon_5^5\ \frac{m_b^2}{m_s}\right]
\ \frac{m_{t}}{m_{c}}
\ \frac{\alpha ^{d^c}}{\alpha^{u^c}\beta ^{d^c}}
-\frac{1}{m_{c}}\ 
\left[m_u+ \varepsilon_5^5\ \frac{m_t^2}{m_c}\right]
\ \frac{1}{\beta^{u^c}}
\right) 
\ .
\label{cabibbo}
\eea
The expression for $(V_{CKM})_{13}$
is written to order \( \varepsilon_5 ^{2} \)
while the other two elements are sufficiently accurate to leading order
in \( \varepsilon_5  \) (note that the mass eigenvalues, written above,
carry powers of $\varepsilon_5$). We therefore have two predictions for the
CKM elements and can eliminate \( \varepsilon_5 \) using the equation for
$(V_{CKM})_{12}$ ;
%
\bea
\label{foep}
\varepsilon_5 = \sqrt{\frac{
(V_{CKM})_{12}
}
{
\left( \frac{m_{t}}{
m_{c}}\ \frac{1}{\alpha ^{u^c}}-\frac{m_{b}}{
m_{s}}\ 
\frac{1}{\alpha ^{d^c}}\right)
}}\ .
\label{epcabibbo}
\eea
%
For instance, 
taking $\alpha^{u^c,d^c}\sim 1$ as discussed in eq.~(\ref{alphabeta}),
and the same numerical example as the one below eq.~(\ref{another3})
we find that
\bea
\varepsilon_5\approx 0.05
\ .
\label{result}
\eea
%
Since $\varepsilon_5=3\ e^{-\frac{2\pi}{3}T_5}$, 
this value corresponds to $T_5\approx 1.95$.
With this result we can check now that
indeed,
as mentioned above, the term proportional to $\varepsilon_5^2$
is negligible in eq.~(\ref{eigenvaluesudapprox2}) but not 
in  eq.~(\ref{eigenvaluesudapprox1}). 

Using the above value for $\varepsilon_5$, and 
$\beta^{u^c,d^c}\sim 1$,
we can also compute now the elements 23 and 13 of the CKM matrix
in eq.~(\ref{cabibbo}) with the result:
\bea
(V_{CKM})_{23}  \approx  0.038 \,\,\, \ , \,\,\,\,\,\,\,
(V_{CKM})_{13}  \approx  0.0026
\ .
\label{V1323}
\eea
%
It is worth noting that the first piece of
$(V_{CKM})_{13}$ in eq.~(\ref{cabibbo}) has the value
0.014, a factor of three too large, but it cancels against the 
the second piece resulting in the correct value.
The values of eq.~(\ref{V1323}) are in quite good agreement
with the experimental ones in eq.~(\ref{ckm}),
considering also that we are assuming
all $\alpha's$ and $\beta's$ equal one,
particular values for quark masses,
and neglecting any renormalization 
effects (eq.~(\ref{quarkmasses}) corresponds to 
Yukawa matrices at the string scale). 
For the sake of comparison, in
addition to the CKM matrix to first order (with a 2$^{nd}$ order 13 element) 
in $\varepsilon_5$ obtained above,
\bea
V_{CKM}=\left( \begin{array}{ccc}
1 & 0.22 & 0.003\\
0.22 & 1 & 0.038\\
0.01 & 0.038 & 1
\end{array}\right) 
\ ,
\label{firstorder}
\eea
we also show it to second order
\bea
V_{CKM}=\left( \begin{array}{ccc}
0.976 & 0.222 & 0.003\\
0.222 & 0.975 & 0.041\\
0.010 & 0.038 & 0.999
\end{array}\right) 
\ ,
\label{secondorder}
\eea
and the one computed numerically
\bea
V_{CKM}=\left( \begin{array}{ccc}
0.978 & 0.210 & 0.002\\
0.210 & 0.978 & 0.039\\
0.010 & 0.038 & 0.999
\end{array}\right) 
\ .
\label{numerically}
\eea
Note that the discrepancy between the numerical and analytic 
values of $(V_{CKM})_{12}$ is because here we are still using the 
first order determination of $\varepsilon_5$ in eq.(\ref{foep}) 
in all the expressions. 

In addition to the magnitudes of the CKM matrix elements we also
require a CP violating phase. 
In string theory, CP violation is problematic because CP is a gauge 
symmetry of the full theory. There has been continued interest in how 
it can be spontaneously broken so that the resulting CP phases are 
physically observable \cite{test,bailin,Giedt,stringcp,ao,ib}. 
Since we are considering entirely renormalizable 
Yukawa couplings, there appear to be only two possibilities here
(in addition to the one already mentioned in footnote 6).
First one can assume that the VEVs of the moduli have an imaginary phase,
which can occur when the flat moduli directions are lifted by 
supersymmetry breaking and find their minimum where the phases are 
non-zero \cite{bailin}. Such a phase feeds directly into 
$\varepsilon_5 $. It is easy to check that this phase is 
physically observable, and leads to a non-zero $\delta $ phase for the 
CKM matrix which is of order one.  An alternative way to break CP 
has been explored in type II D-brane models, which is to break CP without 
breaking supersymmetry by introducing discrete torsion \cite{ao} or Wilson lines 
\cite{ib}. (Torsion would require a factor $Z_N\times Z_M$ orbifold.
To our knowledge CP violation from torsion 
has not yet been examined for the heterotic case.)

Let us conclude this section by using the
mass ratios to fix the relative Higgs VEVs. 
Our results imply that they take the following values;
\bea
v_3^d\approx v_2^d\approx v_1^d
\,\,\,\,\,\,
\ ,
\,\,\,\,\,\,
v_3^u\approx 6 v_2^u\approx 36 v_1^u
\ ,
\label{hiev}
\eea
where we have used eqs.~(\ref{eigenvaluesudapprox1}) and
(\ref{eigenvaluesudapprox2}).
Since the electroweak symmetry breaking condition, 
\bea
(v_1^u)^2 +(v_2^u)^2 +(v_3^u)^2 +(v_1^d)^2 +(v_2^d)^2 +(v_3^d)^2  
= 2\left(\frac{M_W}{g_2}\right)^2
\ , 
\label{ewsb}
\eea
must be fulfilled, using eq.~(\ref{hiev}) 
we obtain 
%
\bea
37 (v_2^u)^2 + 3 (v_2^d)^2 \approx \left(174\ GeV\right)^2
\ . 
\label{ewsb33}
\eea
It is striking that the three $v_i ^d$ VEVs have to be degenerate
to satisfy the experimental values. This is in contrast to the case 
without FI breaking, where the Higgs VEVs had to have the hierarchies observed in
the fermion masses. Here the masses are provided by the hierarchical mixing 
of the physical fields with the FI fields. 
This degeneracy in Higgs VEVs will be advantageous from the 
model building point of view when it comes to the lepton sector, as it 
allows us to permute the elements of the $B^L$ $B^{e^c}$ and $B^\nu $ without
affecting the charged lepton masses. 
This cuts down the number of possibilities for the charged lepton masses, 
thereby increasing predictivity, but allows us more freedom to manipulate the 
neutrino mass matrices, as we now see. 

\section{Lepton mass ratios and mixings}


The mass matrices for leptons before the FI breaking,
given in eq.~(\ref{leptonmasses}),
are as for the quarks, but with different prefactors.
Thus the system is extremely constrained and we will see 
that the correct masses cannot be obtained if we
only include $B^{{e}^c}$ and
$B^{{\nu}^c}$ in the mass matrices.
However, unlike the quarks where 
$B^Q$ 
cannot be present, here we have the possibility of including 
a mixing for the left handed fields as well by including 
$B^L$ in the analysis.
We will see that this improves the results giving the 
correct masses for charged leptons.
For neutrinos this will not be sufficient, but
a see-saw mechanism with a reduced 
see-saw scale breaking will solve the problem.

\subsection{Charged leptons}

Consider first the lepton masses with an FI breaking $B$ matrix for just 
the right handed fields,
$B^{{e}^c}$.
Using eq.~(\ref{newleptonmasses}) we have
\bea
{\cal M}^{e}=gN\epsilon_3a^{{e}^c} A^d B^{e^c}
=gN\epsilon_3a^{{e}^c}
\left( \begin{array}{ccc}
v_1^d \varepsilon_5 \beta^{e^c} & v_3^d\varepsilon_5  & v_2^d\alpha^{e^c} \\
v_3^d\varepsilon_5^2 \beta^{e^c}  & v_2^d & v_1^d\alpha^{e^c} \\
v_2^d\varepsilon_5^2 \beta^{e^c}  & v_1^d\varepsilon_5  & 
v_3^d\alpha^{e^c}/\varepsilon_5
\end{array}\right)
\ .
\label{finaludd}
\eea
%
%
Thus the masses are given by
\bea
\left\{m_e, m_{\mu}, m_{\tau}\right\}
\propto
\left\{v_{1}^d\varepsilon_5 \beta^{e^c},
\, v_{2}^d\ ,\, \frac{v_{3}^d \alpha^{e^c}}{\varepsilon_5 }\right\}
\ .
\label{eigen1}
\eea
%
Since the down-quark masses are determined by eq.~(\ref{eigenvaluesud}),
the ratios are already very constrained. Comparing first to second generation
masses gives
%
\bea
\frac{\beta ^{e^c}}{\beta ^{d^c}}=
\frac{m_{s}}{m_{d}}\
\frac{m_{e}}{m_{\mu }}
\sim 0.05
\ .
\label{ratios}
\eea
%
However,
pattern (\ref{alphabeta}) 
used above in order to obtain the correct quark mass ratios and mixings
implies that $\beta ^{d^c}\sim 1$ and
therefore that $\beta ^{e^c}\sim 0.05\sim \epsilon_5$. The latter
is in contradiction with the three allowed patterns for
$\alpha$'s and $\beta$'s discussed below eq.~(\ref{newleptonmasses2}).
We can try to modify some of the assumptions, by for example
permuting the entries of the $B^{{e}^c}$ with respect to those of $B^{d^c}$,
but in fact this makes the ratios worse, 
and no modification yields charged-leptons masses in a natural way.

Fortunately, as we have already seen, 
the natural situation is for both the left handed and right handed
leptons to combine with other fields. 
In this case we should introduce another matrix,
$B^L$, for
the left handed leptons.
Now using eq.~(\ref{newleptonmasses2})
we obtain
%
%
%
\bea
{\cal M}^{e}=gN\epsilon_3a^La^{{e}^c} B^L A^{d} B^{e^c} 
=gN\epsilon_3a^La^{{e}^c}
\left( \begin{array}{ccc}
v_1^d \varepsilon_5^2 \beta^{L}\beta^{e^c} & 
v_3^d\varepsilon_5  & v_2^d\alpha^{L}\alpha^{e^c}/\epsilon_5 \\
v_3^d\varepsilon_5^3\beta^{L} \beta^{e^c}  & v_2^d & 
v_1^d\alpha^{L}\alpha^{e^c}/\epsilon_5 \\
v_2^d\varepsilon_5^3\beta^{L} \beta^{e^c}  & v_1^d\varepsilon_5  & 
v_3^d\alpha^{L}\alpha^{e^c}/\varepsilon_5^2
\end{array}\right)\ .
\label{finaludddd}
\eea
%
The masses are now given by
\bea
\left\{m_e, m_{\mu}, m_{\tau}\right\}
\propto
\left\{\beta ^{L}\beta ^{e^c}v^{d}_{1}
\varepsilon_5^{2}\ , v^{d}_{2}\ , 
\frac{\alpha ^{L}\alpha ^{e^c}v_{3}^{d}}{\varepsilon_5^{2}}\right\}
\ .
\label{masseselec}
\eea
Concentrating on the electron/down ratios again we now find
\bea
\frac{\beta ^{L}\beta ^{e^c}\varepsilon_5}{\beta ^{d^c}}=
\frac{m_{s}}{m_{d}}
\frac{m_{e}}{m_{\mu }}
\sim 0.05
\ ,
\label{ratiosagain}
\eea
which is remarkably close to the correct value if 
$\beta^{L,e^c}\sim 1$.
Comparing second to third generation masses we also find 
\bea
\frac{\alpha ^{d^c}\varepsilon_5}{\alpha ^{L}\alpha ^{e^c}}=
\frac{m_{\tau }}{m_{\mu }}\ 
\frac{m_{s}}{m_{b}}\sim 1
\ ,
\label{ratiosnew}
\eea
so that we require 
\bea
\alpha ^{L}\alpha ^{e^c}\sim \varepsilon_5 \alpha ^{d^c}
\ .
\label{moreratios}
\eea
If we keep \( \alpha ^{d^c}\sim 1 \) to preserve our good CKM prediction,
this is quite a mild requirement on \( \alpha ^{L} \) and \( \alpha^{e^c} \),
since we just need $\alpha ^{L} \alpha^{e^c}\sim \epsilon_5$.
To satisfy this we need only recall that the two non-trivial patterns
of $B$-matrix have $\alpha\sim 1$, $\beta\sim 1$ or $\alpha\sim
\varepsilon_5$, $\beta\sim 1$. We therefore require that 
$B^L$ and $B^{e^c}$ are of the opposite types.

\subsection{Neutrinos}

\subsubsection{Dirac Neutrino masses with no see-saw}

Turning now to the neutrinos, using eq.~(\ref{newleptonmasses2})
one obtains the following mass matrix;
\bea
{\cal M}^{\nu}=gN\varepsilon_1\varepsilon_3a^La^{{\nu}^c} B^L A^u B^{\nu^c}
=gN\varepsilon_1\varepsilon_3a^La^{{\nu}^c}
\left( \begin{array}{ccc}
v_1^u \varepsilon_5^2 \beta^{L}\beta^{\nu^c} & 
v_3^u\varepsilon_5  & v_2^u\alpha^{L}\alpha^{\nu^c}/\varepsilon_5 \\
v_3^u\varepsilon_5^3\beta^{L} \beta^{\nu^c}  & v_2^u & 
v_1^u\alpha^{L}\alpha^{\nu^c}/\varepsilon_5 \\
v_2^u\varepsilon_5^3\beta^{L} \beta^{\nu^c}  & v_1^u\varepsilon_5  & 
v_3^u\alpha^{L}\alpha^{\nu^c}/\varepsilon_5^2
\end{array}\right)
\ .
\label{finaluddddd}
\eea
%
Thus the neutrino masses are given by
\bea
\left\{m_{\nu_e}, m_{\nu_{\mu}}, m_{\nu_{\tau}}\right\}
\propto
\left\{\beta ^{L}\beta ^{\nu^c} v^{u}_{1}\varepsilon_5^{2}
\ , v^{u}_{2}\ , \frac{\alpha ^{L}\alpha ^{\nu }v_{3}^{u}}
{\varepsilon_5^{2}}\right\}
\ ,
\label{constt}
\eea
and consequently we obtain the ratios
%
\bea
\frac{m_{c}}{m_{u}}
\frac{m_{\nu_e}}{m_{\nu_{\mu} }} &=&
\frac{\beta ^{L}\beta ^{\nu^c}\varepsilon_5}{\beta ^{u^c}}
\sim \varepsilon_5 \ ,
\nonumber \\
\frac{m_{c}}{m_{t}}
\frac{m_{\nu_{\tau}}}{m_{\nu_{\mu} }} &=&
\frac{\alpha ^{L}\alpha ^{\nu^c}}{\alpha ^{u^c}\varepsilon_5}\sim
\frac{\alpha^{\nu^c}}{\alpha^{e^c}} \ ,
\label{ratiosagainn}
\eea
where we have used the results for $\alpha$'s and $\beta$'s derived 
from the quarks and charged leptons. We shall assume that the experimental 
data on neutrino 
mass-squared differences reflects their actual masses; i.e. we 
assume that neutrino masses are hierarchical.

Now consider the second relation in eq.~(\ref{ratiosagainn}). The neutrino 
hierarchy is of the order of 10, and $m_c/m_t \approx 10^{-2}$ 
therefore the 
natural choice would be $\alpha^{\nu^c}\sim \alpha^L\sim 
\varepsilon_5$, $\alpha^{e^c}\sim 1$ 
which determines, $a^{\nu^c}\sim 1$ and $a^L\sim 1$. The heaviest neutrino 
mass is then of order $500$ MeV. Even if the neutrino hierarchies had suggested
$a^{\nu^c}\sim \alpha^L\sim \varepsilon_5$ the largest neutrino 
mass would still have been 1 MeV, about the same as the electron mass.
Again we should remember that the positions of the \( \beta \varepsilon_5  \) 
and \( \alpha /\varepsilon_5  \)
in the mixing matrix \( B^{L} \) were not fixed by the charged leptons 
since the $H_d$ VEVs are degenerate. In other words, because $v^d_i$ are degenerate 
we may permute the positions 
of the $v^u_i$ in the above expressions without changing 
the charged lepton masses. However the largest hierarchy in the $v^u$ is 
$\approx 36$ which is not large enough to bring $m_{\nu_\tau}$ to the required
values.

\subsubsection{Neutrinos masses via the see-saw}

We can solve the above problem through a 
see-saw mechanism \cite{seesaw}, but because the Dirac masses are 
already significantly suppressed, we expect that the required 
see-saw scale will be lower than the usual one. In order 
to avoid introducing any ad-hoc scales into the model
we would ideally like the see-saw scale to be associated with some
other scale already existing in the model. 
At this point we note a striking coincidence; if the Yukawa coupling
for the neutrino is of order \( m_{e} \) and the see-saw scale is
1 TeV, then the expected neutrino mass is
\bea
\frac{m_{e}^{2}}{1\ TeV}=0.25\ eV\ ,
\label{seesawmass}
\eea
which is within an order of magnitude of the experimental values.
This suggests that the most natural situation is one in which a see-saw mass 
of order a few TeV is generated by the electroweak symmetry breaking. Unfortunately 
for the 
examples we examine here, we will find that the see-saw scale is two orders 
of magnitude too large, but 
we think this is an intriguing possibility that is worth pursuing.
The first guess for the neutrino see-saw superpotential is then 
\bea
W^{\nu}\sim  H^u L\nu^c + S\nu^c \nu^c\ ,
\label{seesaw}
\eea
where \( S \) is the same singlet that dynamically generates the
$\mu$ term through the coupling $SH^uH^d$.
(Note in this context that the Giudice--Masiero 
mechanism to generate a $\mu$ term through the
K\"ahler potential is not available for prime orbifolds such as the $Z_3$ 
orbifold \cite{Gava}.)
Therefore $S$ is expected to get a VEV of order 1 TeV.
This is clearly a supersymmetric version of the see-saw mechanism,
but with the new feature that the see-saw scale is tied to the weak
scale\footnote{It is worth pointing out here that the same 
mechanism could be used in the context of the non-supersymmetric
Standard Model, using simply a Majorana mass for the right-handed neutrino
of order 1 TeV.}. 
We can make the hierarchy in the neutrinos less steep by permuting
the mixing to the heavy fields, so that we will be able to generate the 
neutrino masses in the same framework without introducing any hierarchies.

Unfortunately the superpotential is slightly more complicated than 
that in eq.~(\ref{seesaw}) because not all the couplings are allowed.
In fact, for the Z3 model under discussion, the particle assignment 
in eq.~(\ref{assignments}) which can
give a nice suppression to the tau and bottom masses does not allow
the \( S\nu^c \nu^c  \) coupling directly. This is because
the presence of the coupling $SH^uH^d$ implies
that $S$ must be assigned to the following fixed point
components in the first two sublattices:
\bea
S\ \,\,\,\, x\ o
\label{subl}
\eea
On the other hand the coupling $S\nu^c \nu^c$ can only be 
allowed if $S$ is assigned to the same fixed point components as $\nu^c$ 
in the first two sublattices; 
\bea
S\ \,\,\,\, x\ x
\label{subl2}
\eea
However, it is always possible
to couple indirectly through other singlets living at different fixed
points. In that case we can still generate heavy Majorana masses for
the neutrinos, by introducing two more singlets $S'$ and $S''$ 
and modifying the superpotential to
\bea
SH^uH^d  +  SS'S''  +  S'\nu^c\nu^c
\ .
\label{possible}
\eea
This case, where we are using the assignment
\bea
S\ \,\,\,\, x\ o\ \,\,\,\,\,\,\,\,\,\,\,\,
\ S'\ \,\,\,\, x\ x\ \,\,\,\,\,\,\,\,\,\,\,\, \ S''\ \,\,\,\, x\ \cdot
\label{subl3}
\eea
is an obvious generalization of our first guess\footnote{It is possible that
the additional U(1) charges in the string compactification
disallow any such Majorana-like coupling with the same field appearing
twice. In that case other superpotentials with different
assignments are possible, although the phenomenology is somewhat
more complicated. We could have for example
$SH^uH^d  +  SS'\nu^c$ or 
$SH^uH^d  +  S'S''\nu^c$, in which case the right handed neutrino is mixed
with the other singlets, neutral Higgses, winos and binos, 
in a 14$\times $14 or 17$\times $17 neutralino mass matrix.}.
Under this assumption, a canonical Majorana mass for the
right-handed neutrino is generated, and 
the light neutrino mass matrix becomes
\bea
(M^{\nu }) (M^{\nu^c})^{-1}(M^{\nu })^{T}
\label{lightn}
\ ,
\eea
where $M^{\nu }$ is given in eq.~(\ref{finaluddddd})
and $M^{\nu^c}$ arises from the coupling $S'\nu^c\nu^c$
\bea
{\cal M}^{\nu^c}=gNa^{{\nu}^c}a^{{\nu}^c} A^{s'} B^{\nu^c} B^{\nu^c}
\ ,
\label{finaluda}
\eea
with
\bea
A^{s'}=\left( \begin{array}{ccc}
s'_1 & s'_3\varepsilon_5  & s'_2\varepsilon_5 \\
s'_3\varepsilon_5  & s'_2 & s'_1\varepsilon_5 \\
s'_2\varepsilon_5  & s'_1\varepsilon_5  & s'_3
\end{array}\right)
\ .
\label{quarkmasses22}
\eea
%
Note that in eq.~(\ref{lightn}) the mixing \( B^{\nu^c} \)
cancels, so that we may as well assume \( a^{\nu^c} B^{\nu^c }=1 \). 
To present the neutrino mass matrix it is convenient to parameterize
the Higgs VEVs as 
\begin{eqnarray}
v^{u}_{2} & = & c_{l}c_{2}\, \varepsilon_5 ^{1/2}v_{3}^{u}\\
v_{1}^{u} & = & c_{l}c_{1}\, \varepsilon_5 ^{4/3}v_{3}^{u}.
\label{approxVEV}
\end{eqnarray}
Inspection of the numerical values of the VEVs 
in eq.~(\ref{hiev}) shows that the prefactors
\( c_{l}\times c_{1,2} \) are of order unity.
We will first discuss how to obtain a neutrino mass matrix with 
maximal mixing and a hierarchy in the mass eigenvalues of order 10, 
and in the next section will discuss what the appropriate see-saw scale
has to be. We will present two examples of neutrino masses with maximal 
mixing corresponding to the two possible solutions of eq.~(\ref{moreratios}), 
i.e. $\alpha^L, \alpha^{e^c}\sim \varepsilon_5 ,1  $ or 
 $\alpha^L, \alpha^{e^c}\sim 1, \varepsilon_5 $. 
In the next section we shall see that both
possibilities have a see-saw scale of around $10^4$ TeV, and generate an
effective Higgs $\mu$-term of order $500$ TeV. 

First consider choosing \( \alpha^{L}\sim \varepsilon_5  \)
and \( \alpha^{e^c}\sim 1 \) so that $a_L\sim 1$ accordingly. 
In addition we may choose the FI mixing so that 
the $B$-matrix for the leptons is permuted with respect to the quarks.
We recall 
that the degeneracy of the $v_i^d $ VEVs allows us to do this without 
worrying about the charged lepton masses. However, when calculating the 
mixing angles of the MNS matrix, we must be careful to maintain the correct 
$e,\mu,\tau$ generation assignment. (In practice one can take account of this 
by simply using 
the relations in eq.~(\ref{approxVEV}) with the indices permuted, 
but otherwise leaving the generation labels unchanged.)

The mixing is then given by
\bea
B^{L}=\left( \begin{array}{ccc}
1  & 0 & 0\\
0 & 1 & 0\\
0 & 0 & \varepsilon_5 
\end{array}\right)\ .
\label{matrixBB}
\eea
In order to get a form of neutrino matrix that gives maximal mixing (with
nearly 
degenerate entries in the 23 submatrix) we allow the low-energy singlet VEVs
to have a small hierarchy of their own, given by 
\bea
s'_{1} & = & \varepsilon_5^{-3/2}s'_{3}
\nonumber\\
s'_{2} & = & \varepsilon_5^{-1}s'_{3}\ .
\label{labelin}
\eea
To first order in \( \varepsilon_5  \) the mass matrix can then be
written as proportional to
\begin{equation}
\label{nmatrix}
\frac{\varepsilon_5 ^{2}v_{3}^{u2}}{s'_{3}}
\left( \begin{array}{ccc}
0 & 0 & 0\\
0 & 1 & c_{2}\\
0 & c_{2} & c_{2}^{2}
\end{array}\right) +O(\varepsilon_5 ^{5/2}),
\end{equation}
where the 1,2,3 elements correspond to $e,\mu,\tau$ respectively. 
The mass hierarchy between second and third generation is now generically
\( m_{\nu _{3}}/m_{\nu _{2}}\approx \sqrt{\varepsilon_5 } \) which is of 
the right order of 
magnitude. This is because the subleading 
contributions are suppressed by $\varepsilon_5^{1/2}$ with respect 
to the leading terms.
The first generation mass in this case is negligible, and the other
mixing angles (apart from the 23 mixing) can be large, but are sensitive 
to the precise values of parameters such as $c_2$, 
so that it is not possible to get any more predictions in this scheme. 

The second possibility has
\( \alpha^{L}\approx 1  \)
and \( \alpha^{e^c}\approx \varepsilon_5 \) so that 
$a_L=\varepsilon_1, \varepsilon_3, \varepsilon_{1}\varepsilon_{3}$ accordingly. 
In this case we can use 
\bea
B^{L}=\left( \begin{array}{ccc}
\varepsilon_5  & 0 & 0\\
0 & \frac{1}{\varepsilon_5} & 0\\
0 & 0 & 1 
\end{array}\right)\ ,
\label{matrixB2}
\eea
and a very mild singlet hierarchy given by 
\bea
a^{-1}_1 s'_{1}  = a^{-1}_3 s'_3 = \varepsilon_5 s'_{2} \, . 
\label{labelin2}
\eea
where $a_1$ and $a_3$ are of order unity. The neutrino mass
matrix becomes proportional to 
\begin{equation}
\label{nmatrix2}
\frac{v_{3}^{u2}}{s'_{3}}\frac{1}{1-a_1^2a_3^2}
\left( \begin{array}{ccc}
0 & 0 & 0\\
0 & (1-a_1^2a_3^2)c_2^2-a_3^2 & 1\\
0 & 1 & -a_1^2
\end{array}\right) +O(\varepsilon_5 ^{3/2}).
\end{equation}
In this case we again have maximal $\theta_{23}$ mixing, but now
the neutrino mass hierarchy is determined by how close
the constants $a_1$, $a_3$ are to unity. Note that the subleading terms are 
now suppressed by a factor of order $\varepsilon_5^{3/2} \approx 10^{-2} $ 
which implies that $m_{\nu_1}=10^{-2}m_{\nu_3}\sim 5\times 10^{-4}$ eV.

\section{Absolute values }

In the previous sections we were concerned with hierarchies of masses
and with mixings, all of which are independent of any overall prefactors
in the Yukawas and which therefore depended on only one suppression
factor \( \varepsilon _{5} \). We now turn to the absolute values
of Yukawas which do depend on \( \varepsilon _{1,3} \) and also
on additional volume factors.

Having fixed the hierarchies and mixings, we need only concentrate
on one mass eigenvalue of any particular particle, which we choose
to be, $m_{c}$, $m_{s}$, $m_{\mu }$, $m_{\nu _{3}}$,
simply because
these are independent of the various \( \alpha  \) and \( \beta  \)
factors. As discussed earlier, since the right-handed neutrino coupling
is unsuppressed, their values are given by 
\begin{eqnarray}
m_{c} & = & gNa^{u^c}v_{2}^{u}\ ,
\label{various1}
\\ 
m_{s} & = & gN\varepsilon _{1}a^{d^c}v_{2}^{d}\ ,
\label{various2}
\\
m_{\mu } & = & gN\varepsilon _{3}a^{L}a^{e^c}v_{2}^{d}\ ,
\label{various3}
\\
m_{\nu _{3}} & = & 
\left\{
\begin{array}{c}
gN\left(\varepsilon_{1}\varepsilon_{3}\varepsilon_5 \right)^{2}
\frac{(v_{3}^u)^2}{s'_3}\nonumber \\
gN\left(\varepsilon_{1}\varepsilon_{3}a^L\right)^{2}
\frac{(v_{3}^u)^2}{s'_3} \, ,
\end{array}
\right.
\label{various}
\end{eqnarray}
where the two neutrino masses are for the two possibilities outlined in the 
previous section. 
Taking into account eqs.~(\ref{various1}) and (\ref{various2}) above we
can write the constraint (\ref{ewsb33}) as
\bea
\frac{67}{(a^{u^c})^2}
+
\frac{1}{\varepsilon_1^2}\ \frac{0.03}{(a^{d^c})^2}\approx
3\times 10^4
\ , 
\label{ewsb32}
\eea
where we have taken $gN\approx 1$.
On the other hand, following the discussion below eq.~(\ref{af2}), we know that
$a^{u^c,d^c}\sim \varepsilon_1, \varepsilon_3, \varepsilon_1\varepsilon_3$.
Thus, in principle, different combinations between the
values of $a^{u^c}$, $a^{d^c}$ and $\varepsilon_1$ may arise.
A possible solution is obtained taking 
$a^{u^c}\sim a^{d^c}\sim \varepsilon_1\sim \varepsilon_5\sim
0.05$,
implying $v_2^d\approx 40$ GeV and $v_2^u\approx 27$ GeV.
For the ratio
$m_{s}/m_{\mu }$ we then find 
\bea
\frac{\varepsilon _{1}a^{d^c}}
{\varepsilon _{3}a^{L}a^{e^c}}=\frac{m_{s}}{m_{\mu }}\ ,
\label{ration}
\eea
giving
\bea
\varepsilon _{1}\sim \frac{a^{L}a^{e^c}}{a^{d^c}}
{\varepsilon_3}\ .
\label{rationm}
\eea
A solution for this equation is obtained with 
$a^{L}a^{e^c}\sim \varepsilon_1\sim \varepsilon_3$ which is 
consistent with our earlier requirement that 
$\alpha^L\alpha^{e^c}\sim\varepsilon_5$.
Summarizing, we have obtained the result 
$\varepsilon_1\sim \varepsilon_3\sim \varepsilon_5\sim \varepsilon 
\sim 0.05$
implying that $T_1\sim T_3\sim T_5\sim 2$, and
in fact, we could find no other choice of \( a^{u^c,d^c} \)
that would be appropriate.
This last observation is extremely interesting, since
one would naturally expect $T_i$ moduli VEVs that are dynamically 
determined by supersymmetry breaking to be of the same order,
and here we have found that the fermion masses support that.

Concerning the neutrinos, we found 
that mildly hierarchical singlet VEVs led to the correct
neutrino mass hierarchies and mixings with either of the 
possibilities 
\( \alpha^{L}\sim \varepsilon_5\), \( \alpha^{e^c}\sim 1  \),
 \( a^{L}\sim 1 \) or 
\( \alpha^{L}\sim 1\), \( \alpha^{e^c}\sim \varepsilon_5  \),
 \( a^{L}\sim \varepsilon_1,\varepsilon_3 \).
In the first case we find a third generation neutrino
mass of
\bea
m_{\nu _{3}}=gN\varepsilon^{2}_{1}
\varepsilon ^{2}_{3} \varepsilon_{5}^{2}
\frac{v_{3}^{u2}}{s'_{3}}\ ,
\label{capi}
\eea
and taking the ratio of this mass with respect to $m_c$ gives
\bea
\frac{a^{u^c}}{\varepsilon _{1}^{2}
\varepsilon _{3}^{2}\varepsilon ^{2}_{5}}
\frac{v_{2}^{u}s'_{3}}{(v_{3}^{u})^2}=
\frac{m_{c}}{m_{\nu _{3}}}\approx 3\times 10^{10}\ .
\label{capiro}
\eea
which gives us
\bea
s'_{3}\approx 10^4 \ TeV\ ,
\label{escat}
\eea
and therefore, using eq.~(\ref{labelin}),
\bea
s'_{2}\approx 2 \times 10^5\ TeV\,\,\,\,\ , \,\,\,\, s'_{1}\approx 9 \times 10^5 \ TeV
\ .
\label{escato}
\eea
The second neutrino matrix we found gives the same value for $s'_3$ but now we have 
\bea
s'_1\approx s'_3 \approx s'_2/20 \approx 10^4\ TeV \, .
\eea
Summarizing, 
for the neutrinos we found that large mixing angles, and a hierarchy
of \( \sim 10 \) 
are natural if the singlet VEVs
are hierarchical. The experimental values of neutrino masses then
fixed the singlet VEVs (which in this case are acting as a reduced
see-saw scale). The coupling to the Higgs fields,
$SH^uH^d$, is suppressed by
a coupling $\varepsilon$ so that we may assign
an effective \( \mu  \)-term 
values for each generation, the lightest of which is 
(assuming  $\langle S\rangle \sim \langle S'\rangle$)
\bea
\mu _{3}\approx 500 \ TeV \, .
\label{muterm}
\eea
It is not clear (without a full minimization of the potential) how this
would translate into Higgs masses, however we find it remarkable that
they are within reach of the TeV scale required for electroweak symmetry
breaking. One possibility that we will not pursue here, is to try to 
find an assignment 
of fields which allows the fully suppressed neutrino Dirac masses with 
$a^L\approx \varepsilon_5^2$. Alternatively one might try to modify the 
superpotential, with a possible reduction of the $\mu$ term through
a suppressed $SS'S''$ couplings.


\section{Summary of predictions 
with $ \varepsilon _{1}=\varepsilon _{3}=\varepsilon _{5}=\varepsilon $}

All the results of the previous section were used to restrict the
values of \( \varepsilon _{1,3,5} \). However it is interesting
to take the \( \varepsilon _{1}\approx \varepsilon _{3}\approx \varepsilon _{5}=\varepsilon  \)
condition as a starting principle, in order to summarize our 
predictions (see footnote 1).
First the ratios of quark masses, the 12 element 
for the
CKM matrix and \( M_{W} \), fixed \( \varepsilon \approx 0.05 \)
and \( (v_{1}^{u},\, v_{2}^{u},\, v_{3}^{u})\approx (4.5,\, 27,\, 162)
\) GeV
and  \( (v_{1}^{d},\, v_{2}^{d},\, v_{3}^{d})\approx (40,\, 40,\, 40)
\) GeV. It
is interesting that for the downs the VEVs are degenerate. We then
obtained two successful predictions for the CKM matrix elements; 
\begin{eqnarray}
(V_{CKM})_{23}
& = & \frac{1}{\varepsilon }\left(
  \frac{m_{d}}{m_{b}}-\frac{m_{u}}{m_{t}}\right) \\
(V_{CKM})_{13}
& = & \frac{m_{s}}{m_{b}}-\frac{m_{c}}{m_{t}}+\varepsilon \left( \frac{m_{u}}{m_{c}}-\frac{m_{d}}{m_{b}}\frac{m_{t}}{m_{c}}\right) 
\end{eqnarray}
In the charged lepton sector, we found two successful predictions
for mass ratios;

\begin{eqnarray}
\frac{m_{e}}{m_{\mu }} & = & \varepsilon \frac{m_{d}}{m_{s}}\\
\frac{m_{\mu }}{m_{\tau }} & = & \frac{m_{s}}{m_{b}}
\end{eqnarray}
The absolute values of the quark and charged lepton masses gave us
two further predictions;
\begin{eqnarray}
m_{s} & = & \varepsilon m_{c}\\
m_{s} & = & m_{\mu }
\end{eqnarray}
For the neutrinos we found that large mixing angles, and a hierarchy
of \( \sim 10 \) are natural if the singlet VEVs 
are mildly hierarchical. The experimental values of neutrino masses then
fixed the singlet VEVs (which in this case are acting as a reduced
see-saw scale). The coupling to the Higgs fields is suppressed by
a coupling of order $\varepsilon$ so that we may assign
effective \( \mu  \)-term values for each generation, the lightest of which is 
\begin{equation} \mu _{3}\sim 500\ TeV\ .\end{equation}
It is not clear (without a full minimization of the potential) 
whether the $\mu$ terms here can be further reduced (since the singlet 
field in the 
effective $\mu$ term is different from that in the Majorana neutrino
mass 
term, as shown in eq.~(\ref{possible})).
It is also not clear how the mild $s'$ hierarchies would translate into 
$\mu$-term hierarchies and subsequently
into Higgs masses, however we find it intriguing that
this rough estimate of scales gave a result that is within reach of the 
electroweak breaking scale.









\section{Conclusions}

In this paper we have examined the possibility of generating the 
fermion mass structure through purely renormalizable couplings in 
heterotic $Z_3$ orbifolds with two Wilson lines. The advantages of 
these models is that they naturally predict three generations, and also 
that the three generations of Higgs fields give enough freedom to 
allow an entirely geometric explanation of masses and mixings. 
In our analysis we found that the Higgs VEVs required only a mild hierarchy 
in order to fit the experimental values of masses and mixings, and that 
the mass hierarchies were generated by hierarchical mixing with heavy 
fields after FI breaking. This is a central feature of the picture 
presented here. 

Our analysis here has been a phenomenological, `bottom-up', one. That is we 
have assigned the particles to fixed points in a way that can 
reproduce the experimental data complete with hierarchies. 
We leave the search for such an assignment to future work but think that it 
is likely to exist, with issues 
such as anomaly cancellation (by extra triplets and doublets) being 
handled by the string construction. 
In addition we have not completed a full analysis
of minimizing the potential along $D$-flat directions after FI breaking, 
but have made use of the very general features that such a minimization 
should have, namely 
hierarchical mixing of the would-be MSSM fields (i.e. those that couple to the 
Higgs fields) with other doublet and triplet fields that couple to heavy 
fields.

In the sense that our analysis is a `bottom-up' analysis of $Z_3$ orbifolds, 
it is intermediate between full string constructions, and models in less 
restricted extra-dimensional set-ups with fields localized at fixed points 
or on domain walls (see for example ref.~\cite{schmaltz}). The consistency 
conditions of the string constructions provide additional constraints that we
think makes this approach more attractive. Conversely
our approach may prove to be useful in guiding heterotic orbifold model 
building.
Whilst this paper was in preparation, ref.~\cite{ib} appeared. 
In that work a similar 
geometric set-up was presented for type II 
models with intersecting D-branes. 
Although the issue of FI mixing 
did not arise in that case, 
the comparative predictiveness 
is not clear as there are extra `moduli' associated with the 
positions of the D-branes.

\bigskip

\noindent {\bf Acknowledgements}

\noindent 
We gratefully acknowledge O. Lebedev
for his collaboration during the early stages of this work,
mainly in connection with the result of Section~3.1.
We also thanks S. Khalil for interesting discussions.
C. Mu\~noz is grateful to the
members of the IPPP, Durham University, U.K.,
for their kind hospitality, and also for their 
support, through the IPPP Visitor Programme,
to spend the
months of July and August 2002, when most of this work was carried out.
The work of S.A. Abel has been supported by a PPARC Opportunity
grant. 
The work of C. Mu\~noz has been supported 
in part by the CICYT, under contract FPA2000-0980, and
the European Union, under contract 
HPRN-CT-2000-00148.


\begin{thebibliography}{99}

\bibitem{Higgs} P.W. Higgs, `Broken symmetries, massless particles, and
gauge fields', 
{\it Phys. Lett.} {\bf 12} (1964) 132; 

\noindent F. Englert and R. Brout, 
`Broken symmetry and the mass of gauge vector mesons', 
{\it Phys. Rev. Lett.} {\bf 13} (1964) 321; 

\noindent P.W. Higgs, `Broken symmetries and the masses of the gauge
bosons', {\it Phys. Rev. Lett.} {\bf 13} (1964) 508; 

\noindent G.S. Guralnik, C.R. Hagen and 
T.W.B. Kibble, `Global conservation laws and massless
particles', {\it Phys. Rev. Lett.} {\bf 13} (1964) 585; 

\noindent T.W.B. Kibble, `Symmetry breaking in non-abelian gauge
theories', {\it Phys. Rev.} {\bf 155} (1967) 1554. 

\bibitem{Weinberg} S. Weinberg, `A model of leptons', 
{\it Phys. Rev. Lett.} {\bf 19} (1967) 1264; 

\noindent A. Salam, `Weak and electromagnetic interactions',
{\it Proceedings of the Eighth Nobel Symposium}, 1968, ed. N. Svartholm
(Almqvist and Wiksell, Stockholm, 1968; Wiley, New York, 1978), p. 367.


\bibitem{pdg1} K. Hagiwara et al. (Particle Data
Group),
{\it Phys. Rev.} {\bf D66} (2002) 010001.

\bibitem{Cabibbo} N. Cabibbo, `Unitary symmetry and leptonic decays', 
{\it Phys. Rev. Lett.} {\bf 10} (1963) 531; 

\noindent M. Kobayashi and T. Maskawa, 
`CP violation in the renormalizable theory of weak interaction',
{\it Prog. Theor. Phys.} {\bf 49} (1973) 652. 


\bibitem{Torrente} For recent reviews, see for example,
J.W.F. Valle, `Neutrinos: summarizing the state-of-the-art',
hep-ph/0205216;

\noindent P. Aliani, V. Antonelli, R. Ferrari,
M. Picariello and E. Torrente-Lujan,
`The solar neutrino puzzle: present situation and future
scenarios',
hep-ph/0206308;

\noindent S.F. King,
`Neutrino mass models', hep-ph/0208266;

\noindent M.C. Gonzalez-Garcia, `Neutrino masses and mixing: where we 
stand and where we are going', hep-ph/0211054.


\bibitem{mns} Z. Maki, M. Nakagawa and S. Sakata,
`Remarks on the unified model of elementary particles', 
{\it Prog. Theor. Phys.} {\bf 28} (1962) 870.


\bibitem{Dixon} L.J. Dixon, J. Harvey, C. Vafa and E. Witten,
`Strings on orbifolds', 
{\it Nucl. Phys.} {\bf B261} (1985) 678; 
`Strings on orbifolds 2', {\it Nucl. Phys.} {\bf B274} (1986) 285.

\bibitem{Wilson} L.E. Ib\'a\~nez, H.P. Nilles and F. Quevedo, 
`Orbifolds and Wilson lines',
{\it Phys. Lett.} {\bf B187} (1987) 25.

\bibitem{Hamidi}
S. Hamidi and C. Vafa, `Interactions on orbifolds',
{\it Nucl. Phys.} {\bf B279} (1987) 465;

\noindent 
L.J. Dixon, D. Friedan, E. Martinec and S. Shenker,
`The conformal field theory of orbifolds',
{\it Nucl. Phys.} {\bf B282} (1987) 13.

\bibitem{Ib.} L.E. Ib\'a\~nez,
`Hierarchy of quark-lepton masses in orbifolds superstring compactification',
{\it Phys. Lett.} {\bf B181} (1986) 269.

\bibitem{test} J.A. Casas and C. Mu\~noz,
`Fermion masses and mixing angles: a test for string vacua',
{\it Nucl. Phys.} {\bf B332} (1990) 189 [Erratum, ibid. {\bf B340} (1990) 280].

\bibitem{Faustino3} 
J.A. Casas, F. G\'omez and C. Mu\~noz,
`Fitting the quark and lepton masses in string theories',
{\it Phys. Lett.} {\bf B292} (1992) 42.



\bibitem{Faustino} 
J.A. Casas, F. G\'omez and C. Mu\~noz,
`World sheet instanton contribution to $Z_7$ Yukawa couplings',
{\it Phys. Lett.} {\bf B251} (1990) 99.

\bibitem{alemanes} T.T. Burwick, R.K. Kaiser and H.F. M\"uller,
`General Yukawa couplings of strings on $Z_n$ orbifolds'
{\it Nucl. Phys.} {\bf B355} (1991) 689.

\bibitem{japoneses} T. Kobayashi and N. Ohtsubo,
`Geometrical aspects of $Z_n$ orbifold phenomenology'
{\it Int. J. Mod. Phys.} {\bf A9} (1994) 87.

\bibitem{Faustino2} 
J.A. Casas, F. G\'omez and C. Mu\~noz,
`Complete structure of $Z_n$ Yukawa couplings',
{\it Int. J. Mod. Phys.} {\bf A8} (1993) 455.


\bibitem{FayetIliopoulos}
E. Witten, 
`Some properties of $O(32)$ superstrings',
{\it Phys. Lett.} {\bf B149} (1984) 351;

\noindent M. Dine, N. Seiberg and E. Witten, 
`Fayet-Iliopoulos terms in string theory',
{\it Nucl. Phys.} {\bf B289} (1987) 317;

\noindent J.J. Atick, L.J. Dixon and A. Sen,
`String calculation of Fayet-Iliopoulos D terms in arbitrary 
supersymmetric compactifications',
{\it Nucl. Phys.} {\bf B292} (1987) 109;

\noindent M. Dine, I. Ichinose and N. Seiberg,
`F terms and D terms in string theory',
{\it Nucl. Phys.} {\bf B293} (1987) 253.



\bibitem{jhep} C. Mu\~noz,
`A kind of prediction from superstring model building',
{\it JHEP} {\bf 0112} (2001) 015.


\bibitem{Casas2} J.A. Casas and C. Mu\~noz,
`Three generation $SU(3)\times SU(2)\times U(1)_Y$ models from orbifolds',
{\it Phys. Lett.} {\bf B214} (1988) 63.




\bibitem{Kim} L.E. Ib\'a\~nez, J.E. Kim, H.P. Nilles and F. Quevedo,
`Orbifolds compactifications with three families of
$SU(3)\times SU(2)\times U(1)^n$', 
{\it Phys. Lett.} {\bf B191} (1987) 3.

\bibitem{Katehou}
J.A. Casas, E.K. Katehou and C. Mu\~noz, 
`$U(1)$ charges in orbifolds: anomaly cancellation and phenomenological
consequences',
{\it Nucl. Phys.} {\bf B317} (1989) 171.


\bibitem{Font}
A. Font, L.E. Ib\'a\~nez, H.P. Nilles and F. Quevedo, 
`Yukawa couplings in degenerate orbifolds: towards a realistic
$SU(3)\times SU(2)\times U(1)$ superstring',
{\it Phys. Lett.} {\bf B210} (1988) 101.

\bibitem{Casas3} J.A. Casas and C. Mu\~noz,
`Yukawa couplings in $SU(3)\times SU(2)\times U(1)_Y$ orbifolds models',
{\it Phys. Lett.} {\bf B212} (1988) 343.



\bibitem{Paschos} S.L. Glashow and S. Weinberg,
`Natural conservation laws for neutral currents',
{\it Phys. Rev.} {\bf D15} (1977) 1958.


\bibitem{KK} B. McWillians and L.F. Li,
`Virtual effects of Higgs particles',
{\it Nucl. Phys.} {\bf B179} (1981) 62;

\noindent O. Shanker, 
`Flavor violation, scalar particles and leptoquarks',
{\it Nucl. Phys.} {\bf B206} (1982) 253.

\bibitem{Georgi} H. Georgi and D.V. Nanopoulos,
`Supression of flavor changing effects from neutral spinless meson
exchange in gauge theories',
{\it Phys. Lett.} {\bf 82B} (1979) 95.


\bibitem{Cheng} T.P. Cheng and M. Sher,
`Mass matrix ansatz and flavor nonconservation in models with multiple
Higgs doublets',
{\it Phys. Rev.} {\bf D35} (1987) 3484;

\noindent M. Sher and Y. Yuan,
`Rare B decays, rare $\tau$ decays, and grand unification',
{\it Phys. Rev.} {\bf D44} (1991) 1461;

\noindent H.E. Haber and Y. Nir, `Multi-scalar models with a 
high energy scale',
{\it Nucl. Phys.} {\bf B335} (1990) 363.



\bibitem{fcnc} 
N.V. Krasnikov, `Electroweak model with a Higgs democracy',
{\it Phys. Lett.} {\bf B276} (1992) 127;

\noindent A. Antaramian, L.J. Hall and A. Rasin, 
`Flavor-changing interactions mediated by scalars at the weak scale',
{\it Phys. Rev. Lett.} {\bf 69} (1992) 1871;

\noindent A. Aranda and M. Sher,
`Generations of Higgs bosons in supersymmetric models',
{\it Phys. Rev.} {\bf D62} (2000) 092002.




\bibitem{Giedt2} M.K. Gaillard and J. Giedt, 
`A D-moduli problem?', 
{\it Phys. Lett.} {\bf B479} (2000) 308;
`D-moduli stabilization',
hep-ph/0208004;

\noindent T. Barreiro, B. de Carlos, J.A. Casas and
J.M. Moreno, 
`Anomalous U(1), gaugino condensation and supergravity', 
{\it Phys. Lett.} {\bf B445} (1998) 82.

\bibitem{Giedt} J. Giedt, 
`The KM phase in semi-realistic heterotic orbifold models', 
{\it Nucl. Phys.} {\bf B595} (2001) 3 [Erratum-ibid. {\bf B632} (2002) 397];
`CP violation and moduli stabilization in heterotic models',
{\it Mod. Phys. Lett.} {\bf A17} (2002) 1465.


\bibitem{bailin}
B. Acharya, D. Bailin, A. Love, W.A. Sabra and S. Thomas,
`Spontaneous breaking of CP symmetry by orbifold moduli',
{\it Phys. Lett.} {\bf B357} (1995) 387;
 
\noindent D.~Bailin, G.~V.~Kraniotis and A.~Love,
`CP violation by 
soft supersymmetry breaking terms in orbifold  compactifications',
{\it Phys. Lett.} {\bf B414} (1997) 269;
`CP-violating phases in the CKM matrix in orbifolds compactifications',
{\it Phys. Lett.}  {\bf B435} (1998) 323.


\bibitem{stringcp}
S.A. Abel and G. Servant,
`Dilaton stabilization in effective type I string models',
{\it Nucl. Phys.} {\bf B597} (2001) 3;
`CP and flavour in effective type I string models',
{\it Nucl. Phys.}  {\bf B611} (2001) 43;

\noindent O. Lebedev, `The CKM phase in heterotic orbifold models',
{\it Phys. Lett.} {\bf B521} (2001) 71;

\noindent S. Khalil, O. Lebedev and S. Morris,
`CP violation and dilaton stabilization in heterotic string models',
{\it Phys. Rev.}  {\bf D65} (2002) 115014;

\noindent T. Dent,
`Breaking CP and supersymmetry with orbifold moduli dynamics',
{\it Nucl. Phys.}  {\bf B623} (2002) 73
[Erratum-ibid.  {\bf B629} (2002) 493];

\noindent S. Abel, S. Khalil and O. Lebedev,
`The string CP problem',
{\it Phys. Rev. Lett.}  {\bf 89} (2002) 121601;

\noindent A.~E. Faraggi and O. Vives,
`CP violation in 
realistic string models with family universal anomalous  U(1)',
{\it Nucl. Phys.}  {\bf B641} (2002) 93;

\noindent O. Lebedev and S. Morris,
`Towards a realistic picture of CP violation in heterotic string models',
{\it JHEP} {\bf 0208} (2002) 007.

\bibitem{ao} 
S.A. Abel and A.W. Owen,
`CP violation and CKM predictions from discrete torsion', hep-th/0205031.

\bibitem{ib}
D. Cremades, L.E. Ibanez and F. Marchesano,
`Towards a theory of quark masses, mixings and CP-violation', hep-ph/0212064.

\bibitem{seesaw}
M. Gell-Mann, P. Ramond and R. Slansky, `Complex spinors and 
unified theories', in `Supergravity', 
Proceedings of the Workshop, Stony Brook, New York, 1979,
Eds. P. van Nieuwenhuizen 
and
D.Z. Freedman (North-Holland 1979);

\noindent T. Yanagida, in Proceedings of the Workshop on Unified
Theory and Baryon number in the Universe, Tsukuba, Japan, 1979,
Eds. O. Sawada and
A. Sugamoto (KEK 1979);

\noindent R.N. Mohapatra and G. Senjanovic, ``Neutrino mass and
spontaneous parity violation', {\it Phys. Rev. Lett.}
{\bf 44} (1980) 912. 


\bibitem{Gava} G. Lopes-Cardoso, D. L\"ust and T. Mohaupt,
`Moduli spaces and target space duality symmetries in $(0,2)\ Z_N$ orbifold
theories with continuous Wilson lines',
{\it Nucl. Phys.} {\bf B432} (1994) 68;

\noindent I. Antoniadis, E. Gava, K.S. Narain and T.R. Taylor,
`Effective $\mu$-term in superstring theory',
{\it Nucl. Phys.} {\bf B432} (1994) 187.

\bibitem{schmaltz}
E.A. Mirabelli and M. Schmaltz,
`Yukawa hierarchies from split fermions in extra dimensions',
{\it Phys. Rev.}  {\bf D61} (2000) 113011;

G.C. Branco, A. de Gouvea and M.N. Rebelo,
`Split fermions in extra dimensions and CP violation',
{\it Phys. Lett.}  {\bf B506} (2001) 115.



\end{thebibliography}
\end{document}